\documentclass[journal]{IEEEtran}
\usepackage{amsmath}
\usepackage{cite}
\usepackage{amssymb}
\usepackage{amsfonts}
\usepackage{algorithm}
\usepackage{dsfont}
\usepackage{graphicx}
\usepackage{epsfig}
\usepackage{subfigure}
\usepackage{psfrag}
\usepackage{xcolor}
\usepackage{url}
\usepackage[colorlinks,linkcolor=black,urlcolor=black,anchorcolor=black,citecolor=black,hyperfootnotes=true]{hyperref}
\usepackage{tikz}

\usepackage{multirow}
\usepackage{booktabs}
\usepackage{colortbl}

\newtheorem{remark}{Remark}
\newcommand{\mv}[1]{\mbox{\boldmath{$ #1 $}}}

\title{PNC Enabled IIoT: A General Framework for Channel-Coded Asymmetric Physical-Layer Network Coding}
\author{Zhaorui Wang, Ling Liu, Shengli Zhang, Pengpeng Dong, Qing Yang, and Taotao Wang
\thanks{This work of Ling Liu was supported in part by the National Natural Science Foundation of China under Grant 62001300 and in part by the Natural Science Foundation of Guangdong Province of China under Grant 2021A1515011679. The work of Shengli Zhang was supported in part by the Chinese NSF project under Grant 62171291 and in part by the Guangdong Basic and Applied Basic Research Foundation under Grant 2019B1515130003. The work of Qing Yang was supported by the National Natural Science Foundation of China under Grant 61901280. The work of Taotao Wang was supported in part by the Natural Science Fund of Guangdong Province under Grant 2020A1515010708 and in part by the Natural Science Fund of Shenzhen under Grant JCYJ20210324094609027. \emph{(Corresponding author: Shengli Zhang)}}
\thanks{Zhaorui Wang is with the Department of Information Engineering, The Chinese University of Hong Kong, Hong Kong (e-mail: zrwang2009@gmail.com).}
\thanks{Ling Liu is with the College of Computer Science and Software Engineering, Shenzhen University, Shenzhen, China (e-mail: liulingcs@szu.edu.cn).}
\thanks{Shengli Zhang, Qing Yang, and Taotao Wang are with the College of Electronics and Information Engineering, Shenzhen University, Shenzhen, China (e-mails: zsl@szu.edu.cn, yang.qing@szu.edu.cn, ttwang@szu.edu.cn).}
\thanks{Pengpeng Dong is with Huawei Technologies Co., Ltd., Shanghai, China  (e-mail: d47252@huawei.com).}}

\begin{document}
\maketitle \thispagestyle{empty}

\begin{abstract}
This paper investigates the application of physical-layer network coding (PNC) to  Industrial Internet of Things (IIoT) in which a controller and a robot are out of each other's transmission range, and they exchange messages with the assistance of a relay. We particularly focus on a scenario where 1) the controller has more information to transmit than the robot; 2) the channel  of the controller is stronger than that of the robot, and both users have nearly the same transmit power. To reduce the communication latency, we put forth an asymmetric PNC transmission scheme in which the controller transmits more information than the robot by exploiting its stronger channel gain in the uplink of PNC. However, the current channel-coded PNC requires the two users to transmit the same amount of source information in order to preserve the linearity of the two users' channel codes at the relay for successful decoding. Therefore, a challenge in the asymmetric PNC transmission scheme is how to construct a channel decoder at the relay, considering that a superimposed symbol at the relay contains different amounts of source information from the controller and robot. To fill this gap, we propose a lattice-based encoding and decoding scheme in which the robot and  controller encode and modulate their information in lattices with different lattice construction levels. The network-coded messages are decoded level-by-level in the lattice. Our design is versatile on that  the controller and the robot can freely choose their modulation orders based on their channel power, and the design is applicable for arbitrary channel  codes, not just for one particular channel  code. The simulation results demonstrate the effectiveness of the proposed channel-coded asymmetric PNC transmission scheme. 
\end{abstract}

\begin{IEEEkeywords}
Physical layer network coding (PNC), industrial internet of things (IIoT), lattice, channel coding.
\end{IEEEkeywords}

\section{Introduction}\label{sec:Introduction}
\begin{figure}[t]
	\centering
	\includegraphics[scale=0.58]{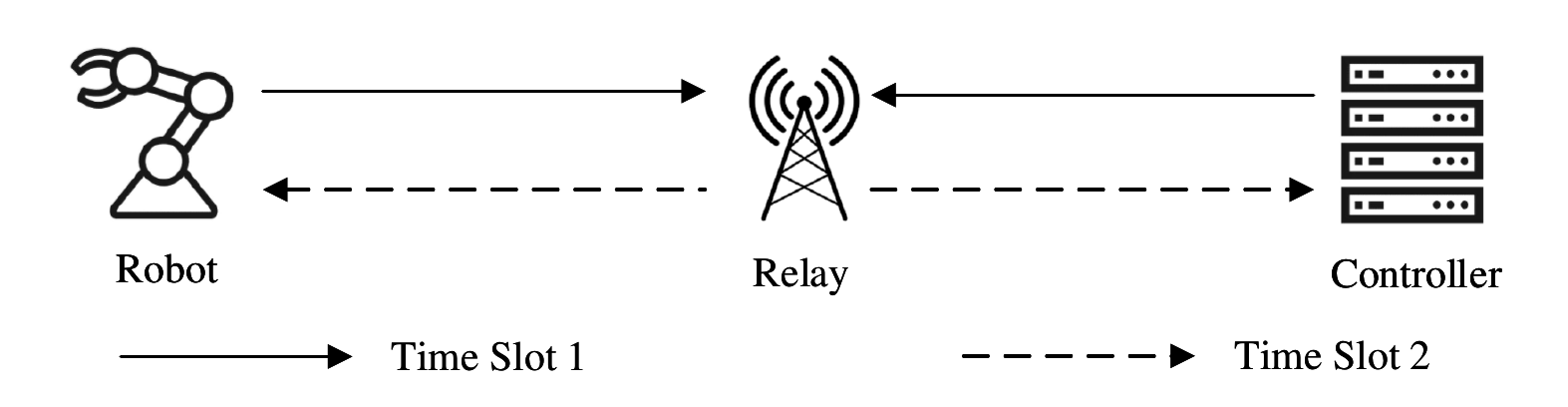}\\
	\caption{A controller and a robot are out of each other's transmission range, and they exchange messages with the assistance of a relay. The PNC technique is applied to reduce the communication latency.}\label{Fig1_1}
\end{figure}

In this paper, we focus on a  scenario in Industrial Internet of Things (IIoT) where a controller and a robot are out of each other's transmission range, and they exchange messages with the assistance of a relay\cite{varga20205g,sisinni2018industrial,yamamoto2018multi,kagawa2017study}. To achieve the stringent requirement on the communication latency between the robot and controller in IIoT, we apply physical layer network coding (PNC)\cite{zhang2006hot,popovski2007physical}, as shown in Fig. \ref{Fig1_1}. Specifically, at time slot 1, the controller and robot transmit their messages simultaneously to the relay. From the overlapped signals, the relay deduces a network-coded message. At time slot 2, the relay  broadcasts the network-coded message to the controller and  robot. The robot then uses the network-coded message and its own message to deduce the message from the controller. Likewise for the controller. Compared with the traditional scheme which requires four times slots for the communications between the robot and controller, PNC can reduce the communication latency from four time slots to two time slots\cite{zhang2006hot,popovski2007physical}.
 
Within the robot and controller communication scenario, we are particularly interested in the case where 1) the message length from the controller is longer than that from the robot. For example, the controller controls motion of the robot through a series of instructions, while the robot only needs to feed back a one-bit acknowledgment to indicate if the robot executes the instructions correctly; 2) the controller and robot have nearly the same transmit power, while the channel power between the controller and relay, is larger than that between the robot and relay. For example, the channel between the controller and relay is a line-of-sight channel, while the channel between the robot and relay is a non-line-of-sight channel due to  the equipment around the robot which creates multipaths and in turn causes channel fading. Another example is, due to the mobility of the robot, the distance between relay and robot could be larger than the distance between relay and controller. In this case, the path loss of the relay-robot channel is larger than that of the relay-controller channel. 

Most of the current channel-coded PNC studies are based on a pioneering work \cite{popovski2007physical}, which showed that as long as the amounts of the information from the two users are the same, the XOR of the two linear codewords from the two users is still a valid codeword at the relay.  Currently, nearly all the PNC channel encoding and decoding techniques were developed over this requirement\cite{zhang2009channel,huang2013design,yang2015achieving,wang2019signal,shi2016subtleties,shi2017complex,wangnoncoherent}, although sometimes  the channel power between the controller and relay is larger than that between the robot and relay. In this case, within our considered scenario, if we apply the  current PNC techniques, the controller should use another time slot to transmit the additional information to the robot separately.

In this paper, to further reduce the communication latency, we put forth an ``asymmetric transmission scheme'' where the controller can transmit more information than the robot  by exploiting its stronger channel gain in the uplink of PNC. In this case, we cannot apply the current PNC channel decoding techniques which require the both users transmit the same amount of information. For example, the robot transmits a QPSK modulated packet. Since the channel between the controller and relay is stronger, we assume that the controller transmits a 8-QAM modulated packet. In addition, the robot and controller apply a same type of channel  code to guarantee the transmission reliability. Since each QPSK symbol contains two encoded bits of the codeword, while each 8-QAM symbol  contains three encoded bits, it is hard for the relay  to find a channel decoder to deduce meaningful network-coded messages from the superimposed packet. Thus, a key challenge is how to construct a channel decoder at the relay to deduce the network-coded messages in the asymmetric transmission.  


\subsection{Related Work}
\underline{\emph{Symmetric transmission with channel codes}}: Currently,  most of the channel-coded PNC studies lie in the symmetric transmission, where two end users transmit the same amount of information, even in the case where the channel gain of the relay-controller channel is larger than that of the relay-robot channel\cite{popovski2007physical,zhang2009channel,huang2013design,yang2015achieving,wang2019signal,shi2016subtleties,shi2017complex,wangnoncoherent}. As long as the amount of the information from the two users is the same, \cite{popovski2007physical} showed that the XOR of the two linear codewords is still a valid codeword at the relay. Thus, the channel decoder can be constructed to deduce the network-coded messages at the relay. However, the problem is that the symmetric transmission does not exploit the larger channel gain from the relay-controller channel such that the controller can transmit more information than the robot. In this case, if the controller has more information to transmit, the controller should use another time slot to transmit the rest of the information.  In this paper, we consider the channel-coded asymmetric PNC  transmission scheme to reduce the transmission time.

\underline{\emph{Asymmetric transmission without channel codes}}: To further exploit the channel gain in the relay-controller channel,  \cite{koike2009adaptive,koike2009optimized,chen2010novel,chen2011optimization,muralidharan2013wireless,chen2013spectrum,zhang2015design} studied the case where the controller transmits more information than the robot,  but without applying channel codes protection. The asymmetric transmission is achieved through a way where the controller chooses a higher signal modulation order than that of the robot\footnote{Since the channel gain of the relay-controller channel is larger than that of the relay-robot channel,  the  received SNR from the controller is larger than that from the robot. For a same target frame error rate (FER), the controller can thus potentially choose a higher order modulation than that of the robot\cite{goldsmith2005wireless}. In this case, the controller can transmit more information with a larger channel gain.}. Refs. \cite{koike2009adaptive,koike2009optimized,chen2010novel,chen2011optimization,muralidharan2013wireless,chen2013spectrum,zhang2015design}  did not apply the channel codes because in the case that the amounts of information from the two users are different, the current PNC coding schemes cannot guarantee the linearity of the underlying channel codes at the relay for successful decoding. In this case, it is not clear how to construct channel decoder at the relay to deduce the network-coded messages. In this paper, we solve this problem by proposing a lattice-based channel encoder and decoder in the asymmetric PNC transmission.

\underline{\emph{Asymmetric transmission with channel codes}}: Prior to this work, \cite{zhang2017design,pan2017practical} put forth novel channel coding and modulation schemes to solve the problem partially in asymmetric PNC transmission. First, the channel coding and modulation scheme in \cite{zhang2017design} is applicable to the case where the robot applies BPSK modulation, and the controller applies QPSK modulation. The channel coding and modulation scheme in \cite{pan2017practical} can be applied to the case where the robot applies $2^m$-QAM modulation, and the controller applies $2^{2m}$-QAM modulation, $m\ge 1$. Second, the channel coding and modulation scheme in \cite{zhang2017design} is particularly designed  for repeat-accumulate (RA)  codes, and channel coding as well as the modulation scheme in \cite{pan2017practical}  is particularly designed for convolutional codes. The detailed description of the encoding and modulation schemes in \cite{zhang2017design,pan2017practical} is shown in Section \ref{sec:related}. In this paper, we put forth a general framework  to solve the above problem comprehensively. Our design is versatile in the following two aspects: 1) in our framework, the controller and robot can freely choose their modulation orders based on their corresponding channel gains; 2) our design is  generally applicable for arbitrary channel codes, not particularly applicable for one type of channel code.

\subsection{Contributions}
First, we put forth a lattice-based channel encoding and modulation framework to solve the channel coding problem in asymmetric PNC. Specifically, a lattice is a discrete set of points in a complex Euclidean space that forms a group under ordinary vector addition\cite{coset1988}. The lattice can be constructed through  a set of nested linear binary channel codes  $\mathcal{C}_1\subseteq\mathcal{C}_2\subseteq\dots\subseteq\mathcal{C}_{L-1}$, where $\mathcal{C}_l$ lies in the $l$-th level of the lattice, $l=1,\dots,L-1$, and $L$ is the number of lattice construction level. That is, the source information is stored in the first $L-1$ lattice levels.  A power shaping in the $L$-th lattice level is applied to constrain the power of the lattice. The lattice with larger construction levels $L$ has larger power.  Denote the number of lattice level at the robot and controller by $L_R$ and $L_C$, respectively. Since the channel at the controller is stronger that that of the robot, we have $L_C>L_R$. The relay estimates the network-coded messages from the received lattice level $l=1$ to lattice level $l=L_C-1$ in a level-by-level manner.

Second, when the lattice levels from the two users are not the same, the conventional power shaping design (i.e., the power shaping design applicable for point-to-point systems) makes the lattices from the two users not nested with each other, since the  power shaping at the robot is not a legal codeword to the channel codes $\mathcal{C}_{L_R}$ in general. Thus, the channel decoder at  the lattice level $L_R$ can not decode network-coded information successfully, which then causes decoding error propagation at the lattice levels $l>L_R$. To solve this problem, we ask the robot to transmit a correction signal beforehand, such that the difference between the power shaping and the correction signal is a legal codeword to the codes $\mathcal{C}_{L_R}$. Upon receiving the superimposed signal, the correction signal is subtracted from the received signal. In this case, the decoder at the relay can estimate the network-coded messages successfully. 

Third, to reduce the correction signal transmission time, we apply the polar source coding \cite{arikan2010source,cronie2010lossless} technique to compress the correction signal, and transmit the compressed correction signal instead. We find that the polar source coding technique can efficiently reduce the correction signal transmission time when the channel coding rate at lattice level $L_R$ is close to 1. We emphasize that this can be achieved when the lattice construction level is large.  In the numerical section, we show this though an example when $L_C=5$. To make the study of the asymmetric transmission comprehensive, we also consider the case where channel coding rate at lattice level $L_R$ is not close to 1. In this case, the length of the compressed correction signal may be large, and the asymmetric transmission scheme may spend  much time on the correction signal transmission in addition to the PNC transmission. Thus, the overall asymmetric transmission time may be larger than the symmetric transmission time. To solve this problem, we put forth a dynamic transmission scheme in which  the relay  dynamically selects one of the transmission schemes which has smaller transmission time.

\subsection{Organization}
The rest of this paper is organized as follows. Section \ref{sec:SYS} describes the system model for the symmetric transmission scheme and asymmetric transmission scheme. In addition, we detail the related work on the channel encoder and modulator in asymmetric transmission scheme in Section \ref{sec:related}. Section \ref{sec:en-de} introduces the
proposed lattice-based channel encoder and modulator in asymmetric transmission, and the power shaping design. Section \ref{sec:nu} presents the numerical results to validate the effectiveness of the proposed asymmetric transmission scheme. Section \ref{sec:dy} proposes a dynamic transmission scheme to solve the problem on which the symmetric transmission time may be smaller than that of  the asymmetric transmission. Finally, Section \ref{sec:con} concludes this paper.

\section{System Model}\label{sec:SYS}
In this paper, we study the communications between a controller and a robot in a two way relay channel (TWRC), as shown in Fig. \ref{Fig1_1}. The controller and robot are out of each other's transmission range, and they exchange messages with the assistance of a relay. In particular, we focus on a scenario in which 1) the message length from the controller is longer than that from the robot; 2) the controller and robot have nearly the same transmit power, while the channel power between the controller and relay, is larger than that between the robot and relay. To  simplify the exposition, we denote the robot by $A$, the controller by $B$, and the relay by $R$. In addition, let $h_u$ be the channel between user $u$ and relay $R$, $u\in\{A,B\}$. From the assumption above, we have  $|h_B|>|h_A|$. We assume that  the coherence time is larger than a packet duration, and thus $h_u$ keeps constant within a packet duration, $u\in\{A,B\}$. Moreover, given the same transmit power, for a same target frame error rate (FER), a channel with stronger power can potentially support a higher modulation order\cite{goldsmith2005wireless}. Suppose the signal modulation order that can be supported by the channel $h_u$ is $M_u$, $u\in\{A,B\}$. In this case, we have $M_B>M_A$. Let $\mv{s}_u\in\{0,1\}^{K_u}$ denote the source information of user $u$, where $K_u$ is the length of the source information, $u\in\{A,B\}$. Under the considered setup, we have $K_B>K_A$. Note that, most of the current PNC studies require that the source information length at both users should be equal to each other, i.e., $K_A=K_B$. To achieve this,  users $A$ and $B$  apply the same coding rate and modulation order\cite{wangnoncoherent,wangcoherent,wangnoncoherent_cof,xie2019polar,yang2014asynchronous,liew2015primer}. To reduce the transmission duration, in this paper, we put forth an asymmetric transmission scheme in which user $B$ can transmit more information than the user $A$, i.e., $K_B>K_A$.

\subsection{Symmetric Transmission Scheme}\label{sec:sy}
We first introduce the conventional symmetric transmission scheme, in which  both users $A$ and $B$ transmit source information with the same length during the PNC phase, and user $B$ transmits the rest of information separately in a point-to-point (P2P) phase.  The signal transmission process is detailed as follows.

\begin{figure}[t]
	\centering
	\includegraphics[scale=0.58]{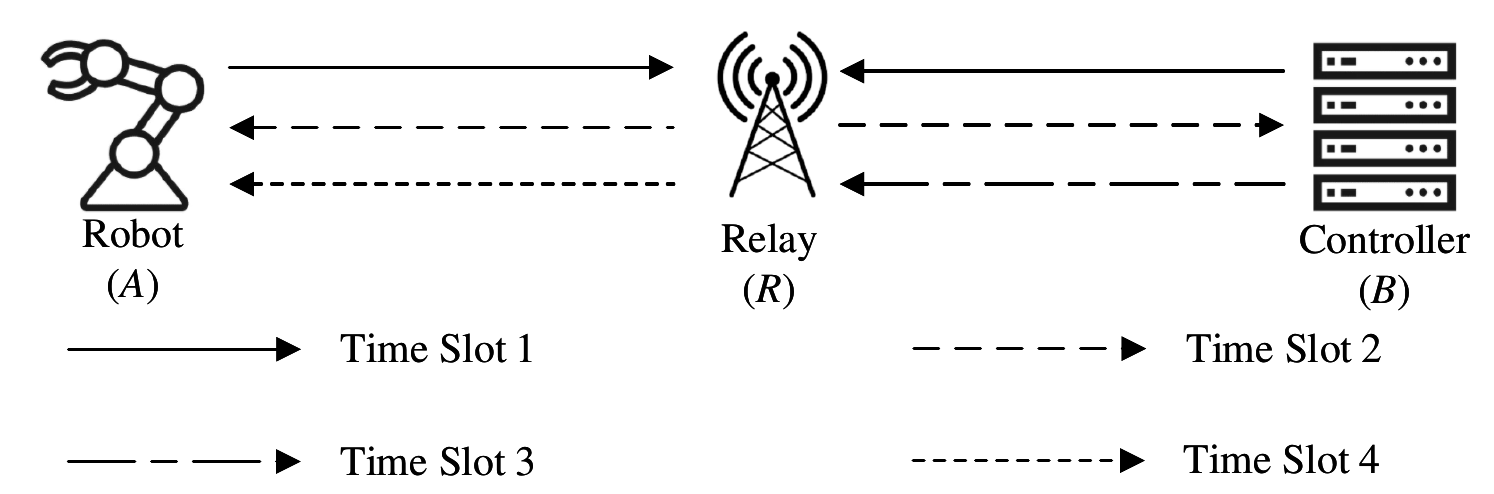}\\
	\caption{Symmetric transmission scheme where the whole transmission  takes four time slots.}\label{Fig1_2}
\end{figure}

\underline{\emph{Time slot 1: Uplink PNC transmission.}} The source information $\mv{s}_B$ from user $B$ is divided into the two parts: $\mv{s}_{B,PNC}$ and $\mv{s}_{B,P2P}$, where $\mv{s}_{B,PNC}\in\{0,1\}^{K_A}$, and $\mv{s}_{B,P2P}\in\{0,1\}^{(K_B-K_A)}$. The information $\mv{s}_{B,PNC}$ is transmitted during the PNC phase, and the information $\mv{s}_{B,P2P}$ is transmitted by user $B$ separately during the P2P phase. In the PNC phase, the source information $\mv{s}_A$ and $\mv{s}_{B,PNC}$ with the same length go through a same channel-encoder-and-modulator (EM), with coding rate $R_{A,PNC}=R_{B,PNC}$, and modulation order $M_A$. Note that the modulation order now is restricted by the weaker channel $h_A$. Otherwise, FER of the uplink transmission will be higher than the target FER. The transmitted packets are $\mv{x}_{A,PNC}$ and $\mv{x}_{B,PNC}$, respectively. We assume that the bandwidth in the uplink and downlink channel is $W$ symbols per second, i.e., the transmitter transmits $W$ modulated symbols to the receiver per second.  The  time slot 1 duration is
\begin{align}
T_1^{(Sym)}=\frac{K_A}{R_{A,PNC}M_AW}.
\end{align}

We assume that the signals from users \emph{A} and \emph{B} arrive at relay \emph{R} simultaneously, the received signal is expressed as:
\begin{align}
&\mv{y}_{R,PNC}\nonumber\\
&=h_A\beta_A\frac{1}{\sqrt{p_A}}\mv{x}_{A,PNC}+h_B\beta_B\frac{1}{\sqrt{p_B}}\mv{x}_{B, PNC}+\mv{n}_{R,PNC},\label{eq:rec}
\end{align}
where $p_u$ is the power of the symbol $x_{u,PNC}^{(n)}$, where $x_{u,PNC}^{(n)}$ is the $n$-th symbol in the packet $\mv{x}_{u,PNC}$, $u\in\{A,B\}$, and $n=1,\dots,N$. In this case, $\frac{1}{\sqrt{p_u}}\mv{x}_{u,PNC}$ denotes a power-normalized packet; $\beta_u$ is the channel precoder to compensate the channel at user $u$, $u\in\{A,B\}$, and $\mv{n}_{R,PNC}\sim\mathcal{CN}\left(\mv{0},\sigma^2_{R,PNC}\mv{I}\right)$ denotes the additive white Gaussian noise (AWGN) at the relay.  In addition, we assume perfect channel precoding at the users, i.e., 
\begin{align}
h_u\beta_u\frac{1}{\sqrt{p_u}}=1, ~~~u\in\{A,B\}.\label{eq:c}
\end{align}
We will show in Section \ref{sec:en-de} that the precoding in \eqref{eq:c} makes a lot of sense in our asymmetric PNC design.  Let us first elaborate more details on \eqref{eq:c}. Specifically, we assume that the channels $h_u$'s are perfectly known at the users $A$ and $B$. Given the channel gain $|h_u|$ and the transmit power (i.e., the precoder power $|\beta_u|^2$), the user $u$ chooses the  signal modulation with order $M_u$ with symbol power $p_u$,  such that
\begin{align}
\sqrt{p_u}=|h_u||\beta_u|, ~~~u\in\{A,B\}.\label{eq:c1}
\end{align}
Thus, for a same $|\beta_u|$, from \eqref{eq:c1} we know that larger channel gain $|h_u|$ can help us to support higher order modulation. Next, the precoder adjusts its phase such that
\begin{align}
\theta_{h_u}+\theta_{\beta_u}=0, ~~~u\in\{A,B\},\label{eq:c2}
\end{align}
where $\theta_{h_u}$ is the phase of the channel $h_u$, and $\theta_{\beta_u}$ is the phase of the precoder $\beta_u$.  The channel precoding technique to achieve  \eqref{eq:c} has been studied and implemented in \cite{tan2018mobile}. Specifically, \cite{tan2018mobile}  mainly solves the three problems: 1) time synchronization between users $A$ and $B$; 2) channel amplitude precoding to achieve \eqref{eq:c1}; 3) channel phase precoding to achieve \eqref{eq:c2}. We refer interested readers to \cite{tan2018mobile} for more details.

Substituting \eqref{eq:c} into \eqref{eq:rec}, we have
\begin{align}
\mv{y}_{R,PNC}=\mv{x}_{A,PNC}+\mv{x}_{B, PNC}+\mv{n}_{R,PNC}.\label{eq:rec_compen}
\end{align}
Based on the received signals $\mv{y}_R$, the relay $R$ deduces network-coded messages from users $A$ and $B$. Note that, since users $A$ and $B$ applies a same EM, the current PNC decoder-and-demodulator\cite{xie2019polar,yang2014asynchronous,liew2015primer} can be applied directly. The estimated network-coded information is denoted by  $\mv{s}_{R,PNC}\in\{0,1\}^{K_A}$.


\underline{\emph{Time slot 2: Downlink PNC transmission.}} The relay then broadcasts the estimated network-coded information to both end users. The relay applies an EM, with coding rate $R_{R,PNC}$ and modulation order $M_A$. The modulation order is restricted by the weaker channel $h_A$ to achieve a target FER for the two users\footnote{In the downlink PNC, the received SNR at the controller is larger than that at the robot since the relay-controller channel is stronger than of the relay-robot channel. To make the FER at the robot and controller both smaller than the target FER, the relay chooses the low order modulation; otherwise, the FER of the robot would be larger than the target FER.}. In addition, for exposition simplicity, we assume that the modulation order in time slot 2 is the same as that in time slot 1, and the downlink PNC can achieve different FERs by adjusting the coding rate $R_{R,PNC}$.  The broadcast packet is $\mv{x}_{R,PNC}$, and the duration of the time slot 2 is
\begin{align}
T_2^{(Sym)}=\frac{K_A}{R_{R,PNC}M_AW}.
\end{align}
At the user $u$, the received signal from the relay is
\begin{align}
\mv{y}_{u,PNC}=\mv{x}_{R,PNC}+\mv{n}_{u,PNC}, u\in\{A,B\}.\label{eq:us}
\end{align}
Note that, the channel $h_u$ has been compensated at user $u$. In this case, $\mv{n}_{u,PNC}\sim\mathcal{CN}\left(\mv{0},\sigma^2_{u,PNC}\mv{I}\right)$ denotes the AWGN at the user $u$ after the channel compensation. The decoder at user $u$ decodes the messages from the other user based on the received signal $\mv{y}_{u,PNC}$ and its own message $\mv{x}_u$. 

\underline{\emph{Time slot 3: User $B$ uplink P2P transmission.}} User $B$ transmits its remaining information  $\mv{s}_{B,P2P}$ with length ($K_B-K_A$). The user $B$ applies an EM with coding rate $R_{B,P2P}$. In addition, since the channel between user $B$ and the relay is stronger than that between user $A$ and the relay, the modulation order now is assumed to be $M_B$. The transmitted packet is $\mv{x}_{B,P2P}$, and the duration of the time slot 3 is 
\begin{align}
T_3^{(Sym)}=\frac{K_B-K_A}{R_{B,P2P}M_BW}.
\end{align}
At the relay $R$, the received signal is
\begin{align}
\mv{y}_{R,P2P}=\mv{x}_{B,P2P}+\mv{n}_{R,P2P}.
\end{align}
Note that, the channel $h_B$ has been compensated at the relay. In this case,  $\mv{n}_{R,P2P}\sim\mathcal{CN}\left(\mv{0},\sigma^2_{R,P2P}\mv{I}\right)$ denotes the AWGN after the channel compensation at the relay.  The decoder at the relay $R$ decodes the messages from  user $B$ based on the received signal $\mv{y}_{R,P2P}$. The estimated information is denoted by  $\mv{s}_{R,P2P}\in\{0,1\}^{(K_B-K_A)}$.

\underline{\emph{Time slot 4: Relay $R$ downlink P2P transmission.}} The relay then transmits the information $\mv{s}_{R,P2P}$ to user $A$. The relay $R$ applies an EM with coding rate $R_{R,P2P}$, and the modulation order $M_A$. The modulation order is restricted by the channel $h_A$. In addition, for exposition simplicity, we assume that the modulation order in time slot 4 is the same as that in time slots 1 and 2, and the downlink P2P can achieve different FERs by adjusting the coding rate $R_{R,P2P}$.  The duration of time slot 4 is 
\begin{align}
T_4^{(Sym)}=\frac{K_B-K_A}{R_{R,P2P}M_AW}.
\end{align}
At the user $A$, the received signal is
\begin{align}
\mv{y}_{A,P2P}=\mv{x}_{R,P2P}+\mv{n}_{A,P2P}.
\end{align}
Note that, the channel $h_A$ has been compensated at user $A$. In this case, $\mv{n}_{A,P2P}\sim\mathcal{CN}\left(\mv{0},\sigma^2_{A,P2P}\mv{I}\right)$ denotes the AWGN after the channel compensation.  The decoder at the user $A$ decodes the messages from  user $B$ based on the received signal $\mv{y}_{A,P2P}$. 

Overall, the transmission time  in the symmetric transmission scheme is
\begin{align}
T^{(Sym)}&=T_1^{(Sym)}+T_2^{(Sym)}+T_3^{(Sym)}+T_4^{(Sym)}\nonumber\\
&=\frac{K_A}{R_{A,PNC}M_AW}+\frac{K_A}{R_{R,PNC}M_AW}\nonumber\\
&~~~+\frac{K_B-K_A}{R_{B,P2P}M_BW}+\frac{K_B-K_A}{R_{R,P2P}M_AW}.\label{eq:time-sy}
\end{align}

\subsection{Asymmetric Transmission Scheme}\label{sec:asy}
A problem in the symmetric transmission scheme is that, in the uplink PNC phase (i.e., time slot 1 in the symmetric transmission), user $B$ transmits signals with a lower modulation order $M_A$, although the channel power between user $B$ and the relay can support user $B$ to transmits signals with a higher modulation order $M_B>M_A$. This takes user $B$ additional time for signal transmission. In this paper, by exploiting the stronger channel at user $B$, we put forth an asymmetric transmission scheme,  in which user $A$ transmits its $K_A$-length source information, and user $B$ transmits its $K_B$-length source information simultaneously during the uplink of  PNC phase. The signal transmission processes are detailed as follows.

\underline{\emph{Time slot 1: Uplink PNC transmission.}} User $A$ transmits it source information $\mv{s}_A$, and user $B$ transmits its source information $\mv{s}_B$ to the relay at the same time. The source information $\mv{s}_u$ goes through an EM with coding rate $R_{u,PNC}$ and modulation order $M_u$, and the transmitted packet is $\mv{x}_{u,PNC}$, $u\in\{A, B\}$. In this case, since the channel from user $B$ is stronger than that of user $A$, we have $M_B>M_A$. We assume that the lengths of source information from the two users are chosen such that the lengths of the transmitted packets from the two users are the same.  The  time slot 1 duration is
\begin{align}
T_1^{(Asy)}=\frac{K_A}{R_{A,PNC}M_AW}=\frac{K_B}{R_{B,PNC}M_BW}.
\end{align}
The received signal at the relay is the same as that shown in \eqref{eq:rec_compen}. A key challenge is how to design an EM at the two users such that the relay can decode the network-coded messages from the two end users. We will show our design on the  encoder and decoder in Section \ref{sec:en-de}. The estimated network-coded information is denoted by  $\mv{s}_{R,PNC}\in\{0,1\}^{K_B}$.

\underline{\emph{Time slot 2: Downlink PNC transmission.}} The relay then broadcasts the estimated network-coded information to the two end users. The relay applies an EM, with coding rate $R_{R,PNC}$ and modulator order $M_A$. The modulator order is restricted by the weaker channel $h_A$ to achieve a targeted FER. In addition, the modulation order is the same as that in time slots 2 and 4 in the symmetric transmission scheme in order to have a fair transmission time comparison later. The broadcast packet is $\mv{x}_{R,PNC}$, and the duration of the time slot 2 is
\begin{align}
T_2^{(Asy)}=\frac{K_B}{R_{R,PNC}M_AW}.
\end{align}
The received signal at the relay is the same as that shown in \eqref{eq:us}. The decoder at user $u$ decodes the messages from the other user based on the received signal $\mv{y}_{u,PNC}$ and its own message $\mv{x}_u$, $u\in\{A,B\}$.  

Overall, the transmission time in the asymmetric transmission scheme is
\begin{align}
T^{(Asy)}&=T_1^{(Asy)}+T_2^{(Asy)}\nonumber\\
&=\frac{K_A}{R_{A,PNC}M_AW}+\frac{K_B}{R_{R,PNC}M_AW}.\label{eq:time-asy}
\end{align}
If we set $R_{R,PNC}=R_{R,P2P}$, from \eqref{eq:time-sy} and \eqref{eq:time-asy}, we have
\begin{align}
	T^{(Sym)}-T^{(Asy)}=T^{(Sym)}_3=\frac{K_B-K_A}{R_{B,P2P}M_BW}. \label{eq:com0}
\end{align}
Eqn. \eqref{eq:com0} shows that the time slot 3 in symmetric transmission scheme is saved by the asymmetric transmission scheme. 

\subsection{Challenge in  Traditional EM in Asymmetric Transmission} \label{sec:tr_scheme}
We first show the traditional EM, i.e.,  the EM applied in the current PNC systems, and its problems when applied to asymmetric transmission scheme through a concrete example. In time slot 1 of the asymmetric transmission scheme shown in Section \ref{sec:asy}, the source information $\mv{s}_u$ first goes through a channel encoder, the  output codeword is $\mv{c}_u$ with codeword length $D_u$, $u\in\{A,B\}$. Suppose that the codeword length $D_B=2D_A$. Then, the codeword $\mv{c}_A$ goes through a BPSK modulator, and $\mv{c}_B$ goes through a QPSK modulator. As a result, the modulated packets $\mv{x}_{A,PNC}$ and $\mv{x}_{B,PNC}$ have the same length. The problem is, since each BPSK symbol within $\mv{x}_{A,PNC}$ contains 1 encoded bit  of the codeword $\mv{c}_A$, while each QPSK symbol within $\mv{x}_{B,PNC}$ contains 2 encoded bits  of the codeword $\mv{c}_B$, it is hard for the relay $R$ to find a channel decoder to deduce meaningful network-coded messages from the superimposed packet between $\mv{x}_{A,PNC}$ and $\mv{x}_{B,PNC}$. In this case, the above traditional  EM is not applicable to the asymmetric transmission in PNC.

\subsection{Related Work on EM in Asymmetric Transmission} \label{sec:related}

Prior to this work, \cite{zhang2017design,pan2017practical} put forth novel schemes to solve the above problem. Specifically, in  \cite{zhang2017design}, user $A$ follows the traditional EM shown in Section \ref{sec:tr_scheme}, and the modulated packet is $\mv{x}_{A,PNC}\in\{-1,1\}^N$, where $N$ is the packet length. To solve the problem detailed in Section \ref{sec:tr_scheme}, user $B$  divides the source information $\mv{s}_B$ into the following two parts: $\mv{s}_{B,1}$ and $\mv{s}_{B,2}$, where  $\mv{s}_{B,i}\in\{0,1\}^{K_{B,i}}$, $i=1, 2$. In \cite{zhang2017design}, $K_{B,1}=K_{B,2}=K_A$. The source information $\mv{s}_{B,1}$  and $\mv{s}_{B,2}$ first go through a same RA channel encoder, and the output codewords are $\mv{c}_{B,1}$ and $\mv{c}_{B,2}$, respectively. Then, the codewords  $\mv{c}_{B,1}$  and $\mv{c}_{B,2}$ go through a BPSK modulator separately, and the output BPSK packets are $\mv{x}_{B,PNC}^I\in\{-1,1\}^N$ and $\mv{x}_{B,PNC}^Q\in\{-1,1\}^N$, respectively. Finally, the QPSK modulated packet  of user $B$ is:
\begin{align}
	\mv{x}_{B,PNC}=\mv{x}_{B,PNC}^I+j\mv{x}_{B,PNC}^Q,
\end{align}
where $j^2=-1$. In this case,  the scheme makes two BPSK modulated packets $\mv{x}_{B,PNC}^I$ and $\mv{x}_{B,PNC}^Q$ embedded in the in-phase and quadrature parts of one QPSK modulated packet $\mv{x}_{B,PNC}$,  respectively. Since each BPSK symbol within $\mv{x}_{A,PNC}$, $\mv{x}_{B,PNC}^I$, and $\mv{x}_{B,PNC}^Q$ all contains 1 encoded bit information of their corresponding  codewords, the traditional channel decoder can be applied to deduce the network-coded messages at the relay. The relay in \cite{zhang2017design} applies a PNC joint channel decoder. Specifically, \cite{zhang2017design} first jointly decodes $\mv{s}_{A}$, $\mv{s}_{B,1}$, and $\mv{s}_{B,2}$ based on the received signal $\mv{y}_{R,PNC}$ in \eqref{eq:rec_compen}. Then, the network-coded message $\mv{s}_{R,PNC}$ is as follows:
\begin{align}
	\mv{s}_{R,PNC}=\left[\mv{s}_A\oplus\mv{s}_{B,1}, \mv{s}_A\oplus\mv{s}_{B,2}\right],\label{eq:net_info}
\end{align}
where $\oplus$ denotes the XOR operation. Ref. \cite{zhang2017design} shows the scheme  where user $A$ applies BPSK modulation, and user $B$ applies QPSK modulation. It is not clear whether the channel encoding and modulation scheme in \cite{zhang2017design} can be extended to the cases beyond BPSK-QPSK combination. The following three factors make the extension difficult:  
\begin{itemize}
	\item \emph{Channel decoder design issue}: The channel decoder at the relay is particularly designed for RA codes. We need to re-design the channel decoder if anther channel code is applied. In addition, the PNC joint channel decoder applied in \cite{zhang2017design} is not widely used due to the decoding complexity issue. Specifically, the decoding complexity increases as the number of input states to the channel decoder. In the above example, there are  $2^{(1+2)}=8$ input states. The number of input states of the scheme exponentially increase with the sum of the modulation orders from the two users, making the joint channel decoder infeasible to high order modulations. 
	\item \emph{PNC XOR mapping issue}: It is not clear how to do PNC XOR mapping beyond the BPSK-QPSK combination.
\end{itemize}

We next introduce the  EM in \cite{pan2017practical}. Specifically, the  EM in \cite{pan2017practical} is the same as that in \cite{zhang2017design} introduced above except that the convolutional code is applied in \cite{pan2017practical}. In addition,   \cite{pan2017practical} applies a  PNC XOR channel decoder. Specifically, \cite{pan2017practical} first applies the PNC XOR mapping between codewords as follows: 
\begin{align}
	\mv{c}_{R,PNC}=\left[\mv{c}_A\oplus\mv{c}_{B,1},  \mv{c}_A\oplus\mv{c}_{B,2}\right].\label{eq:xor}
\end{align}  
Then, the soft information of $\mv{c}_{R,PNC}$ in \eqref{eq:xor} is fed to the channel decoder  to get the network-coded message in \eqref{eq:net_info}. Ref. \cite{pan2017practical} use the same way to deal with the other cases beyond BPSK-QPSK combination. The problems are 1) according to the PNC mapping in \eqref{eq:xor}, the codeword length of user $B$ should always be two times as much as that of user $A$. Thus, the scheme from \cite{pan2017practical} can only be applied to the case where user $A$ applies $2^m$-QAM modulation, and user $B$ applies $2^{2m}$-QAM modulation, $m\ge 1$; 2) The channel decoder at the relay is particularly designed for convolutional codes. We should re-design the channel decoder if anther channel code is applied.

In general, the EM in \cite{zhang2017design,pan2017practical} cannot be generally applied to the cases in which users $A$ and $B$ can freely choose their channel codes, and modulation schemes according to their channel power, e.g., user $A$ applies QPSK modulation, and user $B$ applies 8-QAM modulation with low decoding complexity. In the following, we put forth a lattice-based EM to solve the above problem comprehensively.
\section{Lattice-based EM in Uplink of Asymmetric PNC}\label{sec:en-de}
In Section \ref{sec:asy}, we propose an asymmetric transmission scheme to improve the throughput of the PNC systems. To achieve this, users $A$ and $B$ should apply different coding and modulation strategies such that they can transmit different amount of information in the uplink of PNC. A key challenge is how to design the EM at the two users such that the relay can decode the network-coded messages from the two users. In this section, we propose a lattice-based EM to solve the above problem.

\subsection{Preliminaries for Lattice}\label{sec:pre}
A complex lattice $\Lambda_1$ is a discrete set of points in a complex Euclidean $n$-dimensional space $\mathbb{C}^n$ that forms a group under ordinary complex vector addition, $n\ge 1$\cite{coset1988}. A sublattice $\Lambda_2$ ($\Lambda_2\subseteq\Lambda_1$) induces a partition of $\Lambda_1$ into equivalence groups modulo $\Lambda_2$.  We denote this partition by $\Lambda_1/\Lambda_{2}$. When the number of cosets of $\Lambda_2$ in $\Lambda_1$ is two, the lattice partition is the binary lattice partition. Let $\Lambda_1/\Lambda_2/\dots/\Lambda_{L-1}/\Lambda_L$ denote an $n$-dimensional lattice partition chain for $L\ge 2$. For each partition $\Lambda_l/\Lambda_{l+1}$, a code $\mathcal{C}_l$ over $\Lambda_l/\Lambda_{l+1}$ selects a sequence of coset representatives $a_l\in A_l$, where $A_l$ is a set that contains all the coset representatives of $\Lambda_{l+1}$ in the partition $\Lambda_l/\Lambda_{l+1}$, $1\le l\le L-1$. The construction of the binary lattice requires a set of nested linear binary codes $\mathcal{C}_l$ with codeword length $D$ and source information length $k_l$,  $l=1,\dots, L-1$, and $\mathcal{C}_1\subseteq\mathcal{C}_2\subseteq\dots\subseteq\mathcal{C}_{L-1}$. Let $\pi$ be the natural embedding of $\mathbb{F}^D_2$
into $\mathbb{Z}^D$, where $\mathbb{F}^D_2$ is the binary field. In addition, let $\mv{e}_1$, $\mv{e}_2$,\dots, $\mv{e}_{k_l}$ be a basis of $\mathbb{F}^D_2$ that spans the code $\mathcal{C}_l$. When $n=2$, a vector $\mv{x}$ in the binary lattice is expressed as
\begin{align}
\mv{x}=\sum_{l=1}^{L-1}\phi^{l-1}\sum_{j=1}^{k_l}\alpha_j^{(l)}\pi(\mv{e}_j)+\phi^{L-1}\mv{b},
\end{align} 
where $\phi=1+j$, $\alpha_j^{(l)}\in\{0, 1\}$,  and $\mv{b}\in\mv{G}^D$ with $\mv{G}$ being a set of Gaussian integers. Moreover, the length of $\mv{x}$  now is $N=D$. Furthermore,  if $\mv{x}$ is a baseband transmitted signal, the above lattice construction system combines the channel coding and modulation as a joint process, which is quite different from the traditional EM with separated channel coding and modulation processes. In addition, the power shaping $\mv{b}$ should be carefully chosen such that the transmitted baseband signal $\mv{x}$ is power constrained. We will detail this in Section \ref{sec:pw_sp}.

\begin{figure}[t]
	\centering
	\includegraphics[scale=0.62]{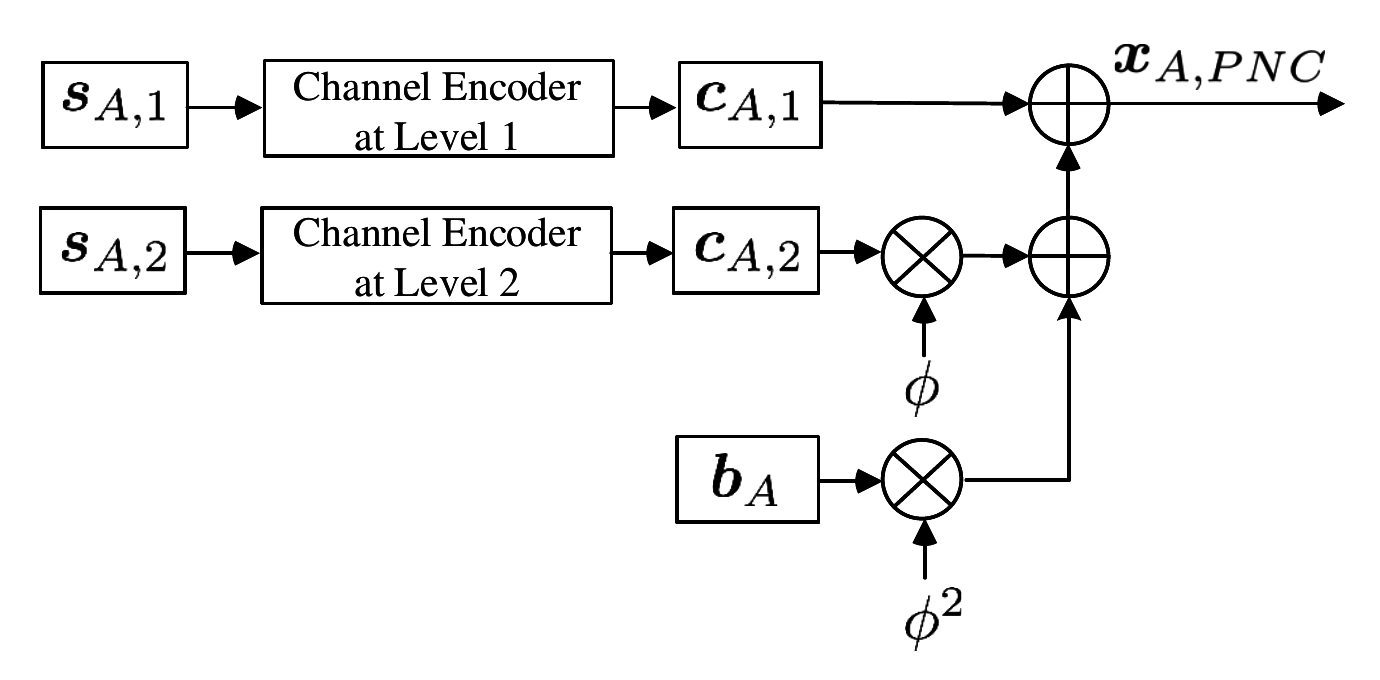}\\
	\caption{A lattice-based EM at user $A$ when $L_A=3$.}\label{Fig2}
\end{figure}
\subsection{Lattice-based Encoder and Decoder}\label{sec:lattice}
Now, we show the lattice constructions at users $A$ and $B$ in the uplink of PNC.  The lattice construction at both users strictly follows the description in Section \ref{sec:pre}. Specifically, according to the channel power, user $u$ applies $L_u$ levels lattice construction, $u\in\{A,B\}$. Then, user $u$ first divides the source information $\mv{s}_u$ as $\mv{s}_u=\left[\mv{s}_{u,1},\dots,\mv{s}_{u,L_u-1}\right]$, where  $\mv{s}_{u,l}$ with length $K_{u,l}$ is the source information at level $l$ in the lattice,  $l=1,\dots, L_u-1$, and $u\in\{A,B\}$. Next, at level $l$, we apply a channel encoder with coding rate $R_{u,l}$ to encode  the source information $\mv{s}_{u,l}$, and the output codeword is $\mv{c}_{u,l}\in\mathbb{Z}^{D_u}$,\footnote{Note that, the codewords $\mv{c}_{u,l}$'s should lie in the whole integer field $\mathbb{Z}^{D_u}$, not in the binary finite field $\mathbb{F}_2^{D_u}$, $u\in\{A,B\}$. The reasons are as follows. In PNC, users $A$ and $B$ transmit signals to the relay simultaneously. Note that, the lattice signals superimposition over the air is actually an addition over the whole integer field. In this case, if we apply binary codewords at the two users, the summation of the two binary codewords from the two users over the air does not lie in the binary finite field anymore, causing the codewords at different lattice levels not nested at the relay. As a result, the decoding failures happen even in the absence of noise at the relay. On the other hand, if we apply codewords that lie in $\mathbb{Z}^{D_u}$, the summation of the two codewords still lie in the whole integer field at the relay. In this case, we can do decoding successfully at the relay.} where $D_u$ is the codeword length, $R_{u,1}\le R_{u,2}\le\dots\le R_{u,L_u-1}$,  $u\in\{A,B\}$, and $l=1,\dots, L_u-1$. Note that, both users should apply a same type of channel  code, e.g., polar codes, LDPC codes, or convolutional codes, during the lattice construction. In addition, the source information length and coding rate at each level of lattice should be the same for the two users, i.e., $K_{A,l}=K_{B,l}$, and $R_{A,l}=R_{B,l}=R_l$, $\forall l$. As a result, we have $D_A=D_B$. Finally, the transmitted packet $\mv{x}_{u,PNC}$, $u\in\{A,B\}$, is expressed as:
\begin{align}
	\mv{x}_{u,PNC}=\mv{c}_{u,1}+\phi\mv{c}_{u,2}+\dots+\phi^{L_u-2}\mv{c}_{u,L_u-1}+\phi^{L_u-1}\mv{b}_{u}. \label{eq:trans}
\end{align}
The length of the transmitted packet $\mv{x}_{u,PNC}$ is $N=D_A=D_B$. We will show how to design the power shaping $\mv{b}_{u}$ later in Section \ref{sec:pw_sp}. In addition, in Fig. \ref{Fig2}, we show an illustrative example of the lattice-based EM at user $A$ when $L_A=3$.

\begin{figure}[t]
	\centering
	\includegraphics[scale=0.58]{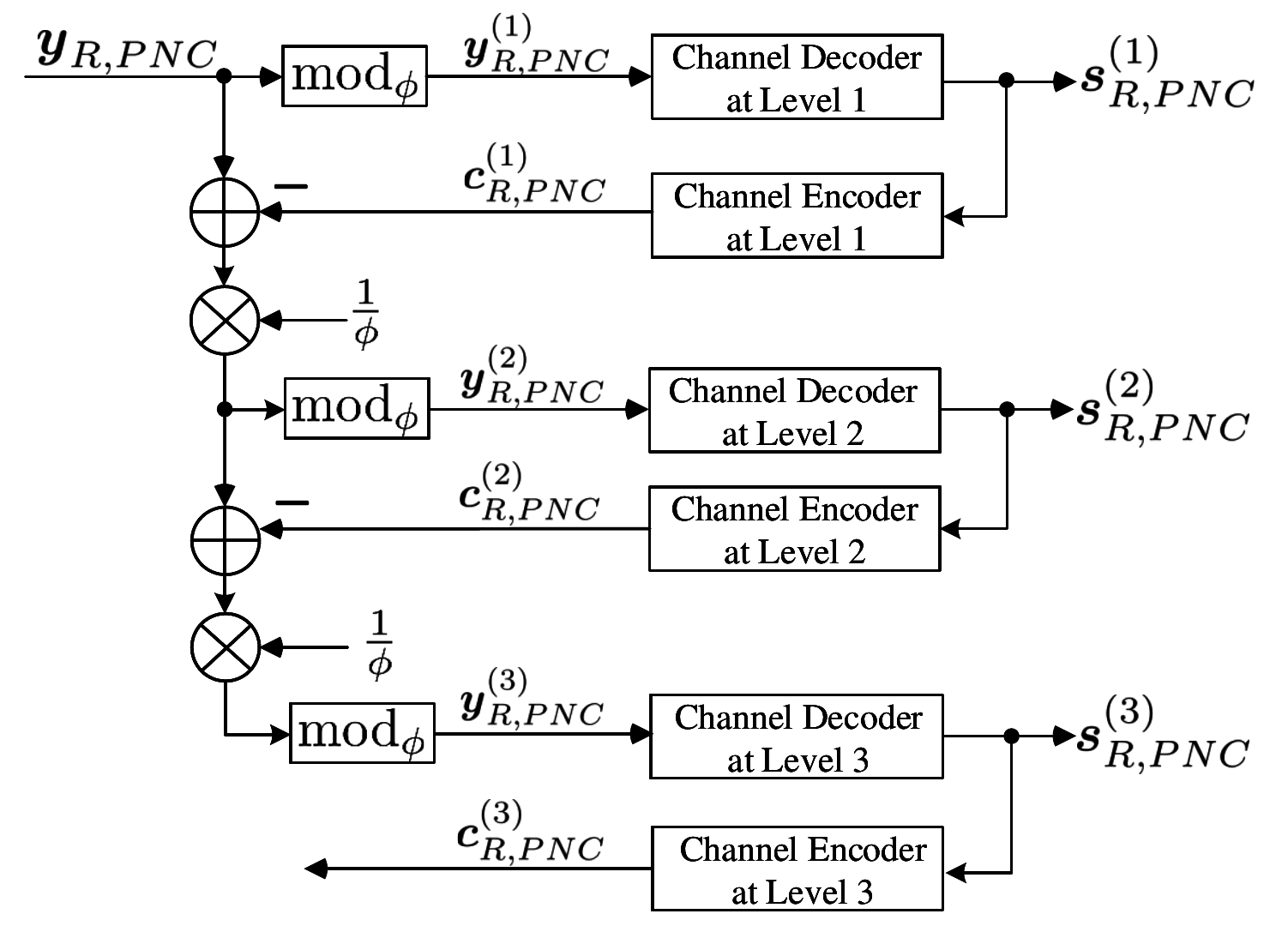}\\
	\caption{A decoder at relay $R$ when $L_A=3$ and $L_B=4$.}\label{Fig3}
\end{figure}

Next, we introduce the decoder at relay $R$. According to \eqref{eq:rec_compen}, the received signal at the relay is
\begin{align}
	&\mv{y}_{R,PNC}\nonumber\\
	&=\mv{x}_{A,PNC}+\mv{x}_{B,PNC}+\mv{n}_{R,PNC}\nonumber\\
	&=(\mv{c}_{A,1}+\mv{c}_{B,1})+\dots+\phi^{L_A-2}(\mv{c}_{A,L_A-1}+\mv{c}_{B,L_B-1})\nonumber\\
	&~~+\phi^{L_A-1}(\mv{b}_A+\mv{c}_{B,L_A})+\dots+\phi^{L_B-1}\mv{b}_{B}+\mv{n}_{R,PNC}.\label{eq:rec_compen2}
\end{align}
From \eqref{eq:rec_compen2}, the effective signals $\mv{x}_{A,PNC}+\mv{x}_{B,PNC}$ forms a $L_B$ levels signal at the relay. The relay decodes the superimposed signals between users $A$ and $B$ level-by-level in the lattice, aiming to deduce the network-coded messages from the two end users. The procedures are summarized as follows:
\begin{itemize}
	\item \textbf{Decode the network-coded message at level 1}.\\
	The decoder at relay $R$  decodes the signal at the first level of the lattice as follows:
	\begin{align}
		\mv{y}_{R,PNC}^{(1)}={\rm mod}_{\phi}(\mv{y}_{R,PNC}),\label{eq:modu}
	\end{align}
	where ${\rm mod}_{\phi}(\mv{y}_{R,PNC})$ denotes $\mv{y}_{R,PNC}$ modulo $\phi$. Through the modulo operation in \eqref{eq:modu}, the resulting signal $\mv{y}_{R,PNC}^{(1)}$ only contains information from the first level of lattice, i.e., ${\rm mod}_{\phi}\left(\mv{c}_{A,1}+\mv{c}_{B,1}\right)$, in which the effective information ${\rm mod}_{\phi}\left(\mv{c}_{A,1}+\mv{c}_{B,1}\right)$ is a BPSK modulated signal. Then, $\mv{y}_{R,PNC}^{(1)}$ is  sent to the channel decoder at the first level of the lattice to estimate the network-coded source information $\mv{s}_{R,PNC}^{(1)}$. Note that the channel decoder should be well-matched to the channel encoder at the each level of the lattice so that the decoding process can be successful. In addition, at each level of lattice, we directly apply the current PNC channel decoders where BPSK modulation is assumed, e.g.,  the LDPC channel decoder, convolutional codes channel decoder, and polar codes channel decoder\cite{xie2019polar,yang2014asynchronous,liew2015primer}. To facilitate the decoding process in the rest levels, $\mv{s}_{R,PNC}^{(1)}$ is re-encoded, and the output codeword is $\mv{c}_{R,PNC}^{(1)}$. 
	\item \textbf{Decode the network-coded message from level 2 to level $L_B-1$ sequentially}.\\ 
	Denote the estimated network-coded source information at level $l$ in the lattice by $\mv{s}_{R,PNC}^{(l)}$, and the corresponding codeword by $\mv{c}_{R,PNC}^{(l)}$, $l=2, \dots, L_B-1$. Then, $\mv{s}_{R,PNC}^{(l)}$ at the level $l$ is computed as follows: 
	\begin{align}
		\mv{y}_{R,PNC}^{(l)}={\rm mod}_{\phi}\left(\hat{\mv{y}}_{R,PNC}^{(l)}\right), l=2, \dots, L_B-1, \label{eq:BAWGN}
	\end{align}
    where
    \begin{align}
    	&\hat{\mv{y}}_{R,PNC}^{(l)}\nonumber\\
    	&=\frac{1}{\phi^{l-1}}\left(\mv{y}_{R,PNC}-\mv{c}_{R,PNC}^{(1)}-\dots-\phi^{l-2}\mv{c}_{R,PNC}^{(l-1)}\right)\nonumber\\
    	&=\frac{1}{\phi^{l-1}}\left(\!\mv{x}_{A,PNC}\!\!+\!\mv{x}_{B,PNC}\!-\!\mv{c}_{R,PNC}^{(1)}\!\!-\!\dots\!\!-\!\phi^{l-2}\mv{c}_{R,PNC}^{(l-1)}\!\right)\nonumber\\
    	&~~~+\frac{1}{\phi^{l-1}}\mv{n}_{R,PNC}\label{eq:noi}.
    \end{align}
	 The $\mv{y}_{R,PNC}^{(l)}$ contains information of ${\rm mod}_{\phi}\left(\mv{c}_{A,l}+\mv{c}_{B,l}\right)$, and is then sent to the channel decoder at the level $l$ of the lattice to estimate the network-coded source information $\mv{s}_{R,PNC}^{(l)}$. Next,  $\mv{s}_{R,PNC}^{(l)}$ is re-encoded through the channel encoder at the level $l$ of the lattice, and the output codeword is $\mv{c}_{R,PNC}^{(l)}$. We compute $\mv{c}_{R,PNC}^{(l)}$ from $l=2$ to $l=L_B-1$ sequentially. In addition,  in \eqref{eq:noi}, since $\frac{1}{\phi^{l-1}}\mv{n}_{R,PNC}\sim\mathcal{CN}\left(\mv{0},\frac{1}{2^{l-1}}\sigma^2_{R,PNC}\mv{I}\right)$, the noise power decreases exponentially as $l$. In this case, through the operation in \eqref{eq:BAWGN}, the channel becomes a binary-input AWGN (BAWGN) channel at level $l$, and the capacity of the BAWGN increases as $l$. Thus, we can transmit much more information at higher levels of the lattice. In particular, when the lattice level $l$ is large, the capacity of the BAWGN at the lattice level $l$ can approach to 1.
\end{itemize}
In Fig. \ref{Fig3}, we show the decoding process at relay $R$ when $L_A=3$ and $L_B=4$ as an illustrative example. Last, the relay encodes the estimated network-coded messages, and broadcasts them to the end users.

\subsection{Power Shaping Design}\label{sec:pw_sp}
In this subsection, we show the power shaping design. To make the every dimension of the transmitted signal $\mv{x}_u$ power constrained, $u\in\{A,B\}$, we apply the hypercube power shaping\cite{tan2018mobile} in our PNC lattice construction in \eqref{eq:trans}. Specifically,  in \eqref{eq:trans}, denote 
\begin{align}
	\mv{c}_u=\mv{c}_{u,1}+\phi\mv{c}_{u,2}+\dots+\phi^{L_u-2}\mv{c}_{u,L_u-1}, u\in\{A,B\}. 
\end{align}
Then, the hypercube power shaping is expressed as follows:
\begin{align}
	\mv{b}_u=\frac{1}{\phi^{L_u-1}}\left({\rm mod}_{\phi^{L_u-1}}\left(\mv{c}_{u}\right)-\mv{c}_{u}\right), u\in\{A,B\}. \label{eq:pw}
\end{align}
In this case, the transmitted packet at user $u$ is
\begin{align}
	\mv{x}_{u,PNC}&=\mv{c}_{u,1}+\phi\mv{c}_{u,2}+\dots+\phi^{L_u-2}\mv{c}_{u,L_u-1}+\phi^{L_u-1}\mv{b}_{u}\nonumber\\
         &={\rm mod}_{\phi^{L_u-1}}\left(\mv{c}_{u}\right), u\in\{A,B\}.
\end{align}
Thus, the power shaping makes the every dimension of the transmitted signal $\mv{x}_{u,PNC}$ power constrained, $u\in\{A,B\}$. For example, when $L_A=3$, $x_{A,PNC}^{(n)}\in\{0, j, -1, -1-j\}$;  when $L_A=4$, $x_{A,PNC}^{(n)}\in\{0, j, -1, -1-j, -j, 1, 1-j, -2j\}$, where $x_{A,PNC}^{(n)}$ is the $n$-th element of $\mv{x}_{A,PNC}$, $n=1,\dots, N$.  

In lattice construction, only hypercube power shaping design shown above can make the transmitted signals power constrained for each dimension. The lattice applying hypercube power shaping works well in point to point communications, and in PNC when $L_A=L_B$. However, in PNC when $L_B>L_A$ is studied in this paper, the hypercube power shaping causes decoding failure at the relay for lattice levels $l\ge L_A$ when ${\rm mod}_{\phi}\left(\mv{b}_A\right)$ is not a codeword to the codebook at the lattice level $L_A$. Specifically, in \eqref{eq:rec_compen2}, at level $L_A$ in the lattice, the signal $\mv{y}_{R,PNC}^{(L_A)}$ contains information of
\begin{align}
	{\rm mod}_{\phi}\left(\mv{b}_A+\mv{c}_{B,L_A}\right)={\rm mod}_{\phi}\left({\rm mod}_{\phi}\left(\mv{b}_A\right)+\mv{c}_{B,L_A}\right). 
\end{align}
Based on $\mv{y}_{R,PNC}^{(L_A)}$, the decoder applies channel decoder to recover the signal ${\rm mod}_{\phi}\left(\mv{b}_A+\mv{c}_{B,L_A}\right)$.  The problem is, ${\rm mod}_{\phi}\left(\mv{b}_A\right)$ may not be a codeword to the codebook at the lattice level $L_A$, i.e., the channel decoder cannot decode ${\rm mod}_{\phi}\left(\mv{b}_A\right)$ successfully even in the absence of noise. When  ${\rm mod}_{\phi}\left(\mv{b}_A\right)$ is not a codeword, the superimposed signal ${\rm mod}_{\phi}\left(\mv{b}_A+\mv{c}_{B,L_A}\right)$ may also not a codeword of the codebook at the lattice level $L_A$. Thus,  the decoder at the level $L_A$ can not recover ${\rm mod}_{\phi}\left(\mv{b}_A+\mv{c}_{B,L_A}\right)$, even in the absence of noise. Moreover, the decoding errors are propagated to the decoders at the lattice levels $l>L_A$. 

To solve the problem, we need to find a way that can not only make the transmitted signal power constrained by applying the power shaping in \eqref{eq:pw}, but also make the decoding in the level $L_A$ successfully. To this end, denote $\mv{c}_{A,L_A}$  a codeword of the codebook at the lattice level $L_A$. In this case, we propose to ask user $A$ to transmit a correction signal $\mv{e}\in\{0,1\}^N$ to the relay beforehand such that
\begin{align}
	{\rm mod}_{\phi}\left(\mv{c}_{A,L_A}\right)={\rm mod}_{\phi}\left(\mv{b}_A-\mv{e}\right).\label{eq:bsc}
\end{align}
Then, users $A$ and $B$ transmit their signals to the relay simultaneously.  Upon receiving the superimposed signals $\mv{y}_{R,PNC}$ as shown in \eqref{eq:rec_compen2}, the correction signal $\mv{e}\in\{0,1\}^N$ is subtracted from $\mv{y}_{R,PNC}$, and the resulting signal is
\begin{align}
	&\hat{\mv{y}}_{R,PNC}\nonumber\\
	&=\mv{y}_{R,PNC}-\phi^{L_A-1}\mv{e}\nonumber\\
	&=\mv{x}_{A,PNC}+\mv{x}_{B,PNC}-\phi^{L_A-1}\mv{e}+\mv{n}_{R,PNC}\nonumber\\
	&=(\mv{c}_{A,1}+\mv{c}_{B,1})+\dots+\phi^{L_A-2}(\mv{c}_{A,L_A-1}+\mv{c}_{B,L_A-1})\nonumber\\
	&~~~~+\phi^{L_A-1}(\mv{b}_{A}-\mv{e}+\mv{c}_{B,L_A})+\dots+\phi^{L_B-1}\mv{b}_{B}+\mv{n}_{R,PNC}.\label{eq:rec_compen3}
\end{align}
In this case, at the level $L_A$, according to \eqref{eq:bsc}, since
\begin{align}
{\rm mod}_{\phi}\left(\mv{b}_{A}-\mv{e}+\mv{c}_{B,L_A}\right)={\rm mod}_{\phi}\left(\mv{c}_{A,L_A}+\mv{c}_{B,L_A}\right)
\end{align}  
is a codeword to the codebook at the lattice level $L_A$, the decoder can recover ${\rm mod}_{\phi}\left(\mv{c}_{A,L_A}+\mv{c}_{B,L_A}\right)$ successfully through the decoding and encoding process. In this case, the transmitted signals at the two users are power constrained, and the relay can decode the superimposed signals successfully. There are two problems to be solved:

\underline{\emph{A: Find a codeword around the hypercube power shaping.}} From \eqref{eq:bsc}, we have
\begin{align}
{\rm mod}_{\phi}\left(\mv{b}_A\right)={\rm mod}_{\phi}\left({\rm mod}_{\phi}\left(\mv{c}_{A,L_A}\right)+\mv{e}\right).\label{eq:bsc1.1}
\end{align}
We can imagine that the codeword ${\rm mod}_{\phi}\left(\mv{c}_{A,L_A}\right)$ goes through a binary symmetric channel (BSC) with bit flipping probability $p$, the correction signal $\mv{e}$ is the corresponding BSC noise, and ${\rm mod}_{\phi}\left(\mv{b}_A\right)$ is the output of the BSC channel.  The capacity $C_{BSC}$ of the BSC is\cite{cover1999elements}
\begin{align}
	C_{BSC}=1-H(p), \label{eq:bscc}
\end{align}
where
\begin{align}
	H(p)=-p\log_2(p)-(1-p)\log_2(1-p),\label{eq:entro}
\end{align}
is binary entropy function. To find the codeword $\mv{c}_{A,L_A}$, we send ${\rm mod}_{\phi}\left(\mv{b}_A\right)$ to the BSC channel decoder with the output source information  $\mv{s}_{A,L_A}$. The BSC channel decoder is similar to the channel decoder at the lattice level $L_A$, and the only difference is that the channel decoder is constructed under binary AWGN channel, while the BSC channel decoder is constructed under BSC channel. The source information $\mv{s}_{A,L_A}$ is then re-encoded, with output codeword $\mv{c}_{A,L_A}$. The correction signal (i.e., the BSC noise) $\mv{e}$ is
\begin{align}
	\mv{e}={\rm mod}_{\phi}\left(\mv{b}_A-\mv{c}_{A,L_A}\right). \label{eq:bsc2}
\end{align}

The above decoding process, i.e., finding ${\rm mod}_{\phi}\left(\mv{c}_{A,L_A}\right)$ from ${\rm mod}_{\phi}\left(\mv{b}_A\right)$ with $\mv{e}$ being the noise vector, can be interpreted as a lossy compression process. We denote the space ${\rm mod}_{\phi}\left(\mv{c}_{A,L_A}\right)$ lies in by $\{0,1\}^{K_{L_A}}$, i.e., the dimension of the space is $K_{L_A}$ although the length of the codeword $\mv{c}_{A,L_A}$ is $N$, and the space ${\rm mod}_{\phi}\left(\mv{b}_A\right)$ lies in by $\{0,1\}^{N}$. In this case, the BSC decoding process actually compresses the space $\{0,1\}^{N}$ to the space $\{0,1\}^{K_{L_A}}$.  Ref. \cite{korada2010polar} proves that, as $N$ goes to infinity, for the lossy compression under the measure of Hamming distortion, the optimal compression rate can be expressed as
\begin{align}
\frac{K_{L_A}}{N}=R_{L_A}=1-H(p), \label{eq:Compre_rate}
\end{align}
where $H(p)$ is the entropy of $\mv{e}$. In this case, we can determine the flipping probability according to 
\eqref{eq:Compre_rate}.

\begin{figure}[t]
	\centering
	\includegraphics[scale=0.62]{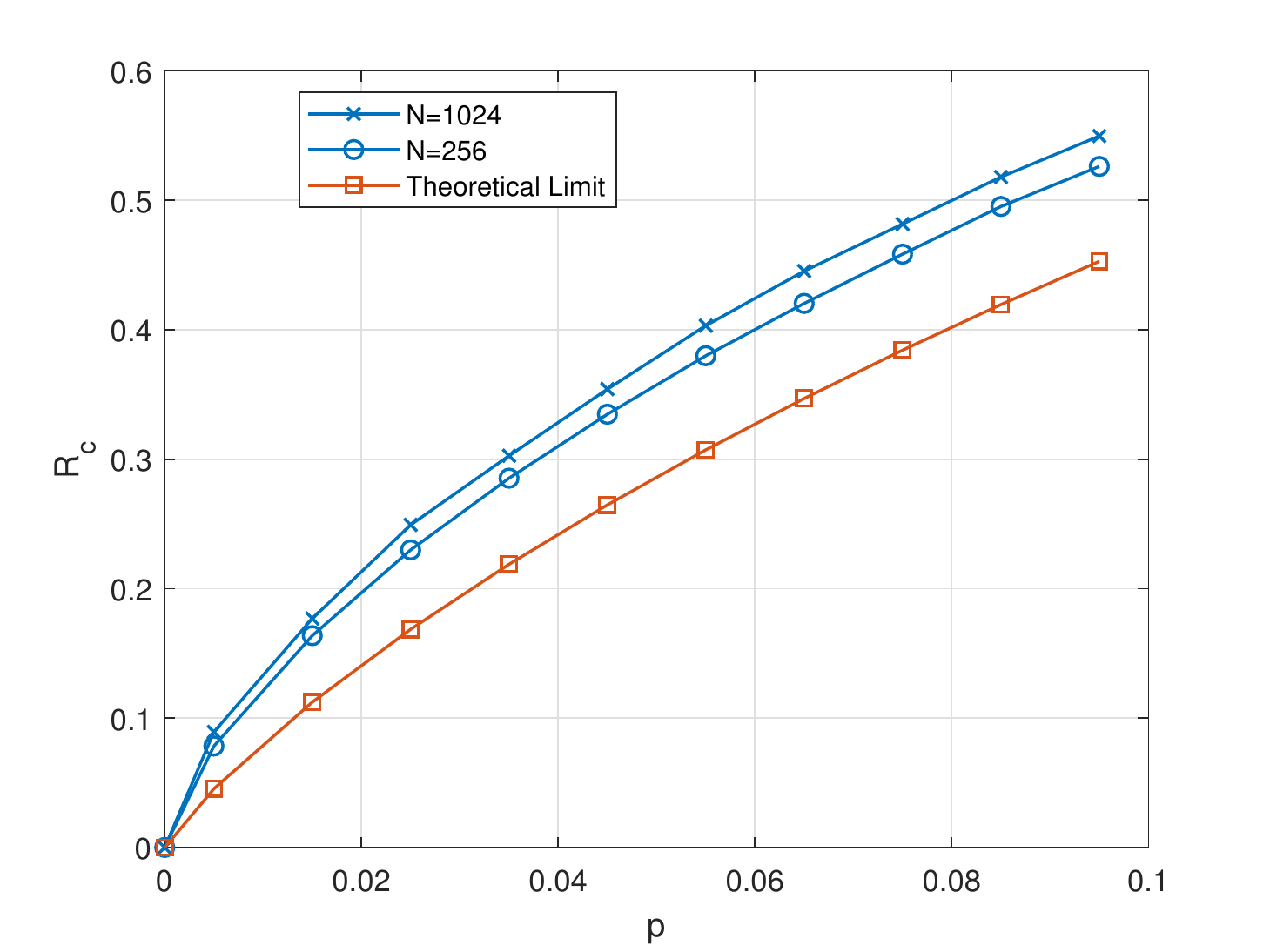}\\
	\caption{Compression rate $R_c$ over flipping probability $p$ for different $N$. }\label{Fig4}
\end{figure}

\underline{\emph{B: Compress signal in \eqref{eq:bsc2} by polar source coding.}} To facilitate the decoding process in the relay as shown in \eqref{eq:rec_compen3}, we need to transmit the correction signal $\mv{e}$ in \eqref{eq:bsc2} to the relay. However, if we ask the user $A$ to transmit the correction signal $\mv{e}$ with length $K_{e}=N$ directly, the time saved through the asymmetric PNC transmission will be canceled out by the correction signal transmission. Fortunately, according to \eqref{eq:Compre_rate}, if the rate $R_{L_A}$ is large, then the entropy of  $\mv{e}$ would be small.  In this case, we can apply the lossless polar source coding\cite{arikan2010source} to compress the correction signal $\mv{e}$. Specifically, according to \cite{arikan2010source}, as $N$ goes to infinity, the compression rate  is  $H(p)$ shown in \eqref{eq:entro}. In this case, we can  transmit the compressed correction signal $\hat{\mv{e}}$ with the length $K_{\hat{\mv{e}}}=NH(p)$ instead of the original correction signal $\mv{e}$ with length $K_{\mv{e}}=N$, greatly saving the correction signal transmission time. In practice, for $N$ is finite, we apply the lossless polar source coding  algorithm proposed in \cite{cronie2010lossless}. According to the algorithm, we can perfectly recover $\mv{e}$ from $\hat{\mv{e}}$, and $\mv{e}$ is plugged into \eqref{eq:rec_compen3} for PNC decoding. The algorithm details are omitted. 

In Fig. \ref{Fig4}, we show the compression rate
\begin{align}
	R_c=\frac{K_{\hat{\mv{e}}}}{N}
\end{align}
The theoretical limit is $R_c=H(p)$ shown in\cite{arikan2010source}, and the other two lines show the results  in \cite{cronie2010lossless} for packet length  $N=256$ and $N=1024$. Fig. \ref{Fig4} shows that, when the flipping probability is small, we can reduce the length of the correction signal $\mv{e}$ significantly by applying the lossless polar source coding. In addition, as the packet length $N$ becomes large, the compression rate in \cite{cronie2010lossless} approaches to the theoretical limit.

\begin{figure}[t]
	\centering
	\includegraphics[scale=0.6]{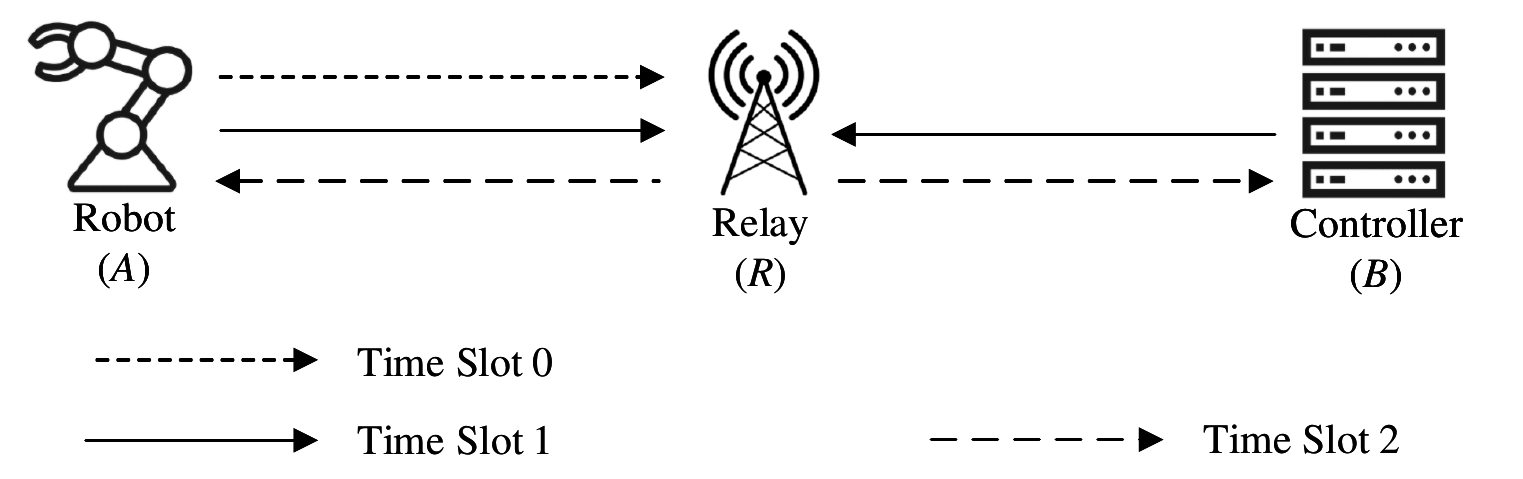}\\
	\caption{Overall asymmetric transmission scheme with correction signal transmission taken into account.}\label{Fig1_3}
\end{figure}

\subsection{Overall Asymmetric Transmission Scheme}\label{sec:overall}
The overall asymmetric transmission scheme is shown in Fig. \ref{Fig1_3}. The signal transmission processes are detailed as follows.

\underline{\emph{Time slot 0: Compressed correction signal $\hat{\mv{e}}$ transmission.}} The user $A$ first computes the correction signal $\mv{e}$ according to the method introduced in Section \ref{sec:pw_sp}, then compresses the correction signal by applying lossless polar source coding according to the algorithm in \cite{cronie2010lossless}. Next, user $A$ transmits the compressed correction signal to the relay. We assume that the user $A$ applies an EM with coding rate $R_{A,P2P}$ and modulator order $M_A$ to transmit the source information $\hat{\mv{e}}$. The modulator order is restricted by the weaker channel $h_A$ to achieve a targeted FER. The transmit packet is  $\mv{x}_{A,P2P}$, and duration of the correction signal transmission is 
\begin{align}
T_{\hat{\mv{e}}}=\frac{K_{\hat{\mv{e}}}}{R_{A,P2P}M_AW}.
\end{align}
At the relay $R$, the received signal is
\begin{align}
	\mv{y}_{R,P2P}=\mv{x}_{A,P2P}+\mv{n}_{R,P2P},
\end{align}
where $\mv{n}_{R,P2P}\sim\mathcal{CN}\left(\mv{0},\sigma^2_{R,P2P}\mv{I}\right)$ denotes the additive white Gaussian noise (AWGN) at the relay in the P2P phase. Note that, the channel $h_A$ has been compensated at the relay. The decoder at the relay $R$ first decodes the compressed correction signal $\hat{\mv{e}}$, and then decompressed it to recover the original correction signal $\mv{e}$ by applying the algorithm in \cite{cronie2010lossless}.

\underline{\emph{Time slot 1: Uplink PNC transmission.}}
The encoding and  modulation processes are the same as that introduced in Section \ref{sec:asy}. To solve the problem that the users $A$ and $B$ can transmit different  amount of information, we apply the lattice-based EM  shown in Section \ref{sec:lattice}. In addition, to make the decoding at the relay successfully, the estimated correction signal is subtracted from the received signal, and the resulting received signal is shown in \eqref{eq:rec_compen3}. Moreover, we apply the lattice-based channel-decoder-and-demodulator to estimate the network-coded messages shown in \ref{sec:lattice}. 

\underline{\emph{Time slot 2: Downlink PNC transmission.}} The encoding and  decoding processes are the same as that introduced in Section \ref{sec:asy}.

In this case, the total transmission time of the asymmetric transmission scheme in \eqref{eq:time-asy} is rewritten as
\begin{align}
&\hat{T}^{(Asy)}\nonumber\\
&=T_1^{(Asy)}+T_2^{(Asy)}+T_{\hat{\mv{e}}}\nonumber\\
&=\frac{K_A}{R_{A,PNC}M_AW}+\frac{K_B}{R_{R,PNC}M_AW}+\frac{K_{\hat{\mv{e}}}}{R_{A,P2P}M_AW}.\label{eq:time-asy2}
\end{align}
If we set $R_{R,PNC}=R_{R,P2P}$, from \eqref{eq:time-sy} and \eqref{eq:time-asy2}, we have
\begin{align}
	T^{(Sym)}-\hat{T}^{(Asy)}&=T^{(Sym)}_3-T_{\hat{\mv{e}}}\nonumber\\
	&=\frac{K_B-K_A}{R_{B,P2P}M_BW}-\frac{K_{\hat{\mv{e}}}}{R_{A,P2P}M_AW}. \label{eq:com}
\end{align}

\begin{remark}\label{R1}
As we discussed in Section \ref{sec:lattice}, as the increase of the lattice level $l$, the capacity of the BAWGN increases even approaching to 1. In this case, the coding rate $R_{L_A}$ can approach to 1 potentially.  Then, according to \eqref{eq:Compre_rate}, we know that $H(p)$ approaches to zero as the increase of the lattice level. In this case,  we can compress the signal $\mv{e}$ to $\hat{\mv{e}}$ with length $K_{\hat{\mv{e}}}=NH(p)$ approaching to zero when $L_A$ is large enough, and the correction signal transmission time $T_{\hat{\mv{e}}}$ can be far smaller than $T^{(Sym)}_3$. In particular, if the coding rate $R_{L_A}=1$, we have $T_{\hat{\mv{e}}}=0$.  We  show this case though the example when user $A$ applies 3-order modulation, and user $B$ applies 4-order modulation in Section \ref{sec:nu}. Moreover, to make the study of the asymmetric comprehensive, we also consider the case when the coding rate $R_{L_A}$ is not close to 1. In this case, the length of the compressed correction signal $\hat{\mv{e}}$ would be large, and may leading to the transmission time $T_{\hat{\mv{e}}}$ larger than $T^{(Sym)}_3$. In this case, the throughput of the proposed asymmetric transmission scheme may be smaller than that of the symmetric transmission. We solve this problem in Section \ref{sec:dy}.
\end{remark}

\subsection{ Discussions on the Lattice Decoder in Section \ref{sec:lattice}}
In the following, we briefly discuss some theoretical analysis on the lattice decoder shown in Section \ref{sec:lattice}. First, we discuss the decoding complexity. Let $D$ denote the overall decoding complexity of the lattice decoder shown in Section \ref{sec:lattice}, and $D_l$ denote the decoding complexity at the $l$-th lattice level, $l=1,\dots, L_{B}-1$. Since the decoding is implemented layer-by-layer,  the overall decoding complexity is
	\begin{align}
		D=D_1+\dots+D_{L_B-1}.
	\end{align}
When we apply polar codes to the lattice-based encoder, since the decoding complexity of the polar codes at the $l$-th level of the lattice is $D_l=\mathcal{O}\left(N\log_2N\right)$, $l=1,\dots, L_B-1$\cite{arikan2009channel}, the overall decoding complexity is $D=\mathcal{O}\left((L_B-1)N\log_2N\right)$. Similarly, we can also compute the decoding complexity when we apply other channel codes to the lattice-based encoder and decoder. We omit the details here.

Second, we discuss the decoding properties of the lattice decoder at the relay. As we show in Section \ref{sec:overall}, the relay begins to decode the PNC signals after the decoding of the correction signals $\mv{e}$. In the case the correction signal $\mv{e}$ is correctly decoded at the relay, the effective received signal $\mv{x}_{A,PNC}+\mv{x}_{B,PNC}-\phi^{L_A-1}\mv{e}$ forms a lattice, whose decoding process is similar to that in the point-to-point system, except that the decoder output in PNC is the network-coded messages.  In this case, we can study the properties of the decoder by applying the tools in lattice. Currently, there are lots of studies on the decoding properties of the lattice in a point-to-point system. For example, the upper bound of decoding FER of the lattice was shown in \cite{liu2018construction} when applying polar codes; the coding gain of the lattice, and the trade-off between the coding and FER was shown in  \cite{forney2000sphere}. These properties can be  applied to the PNC systems. However, in general, the correction signal $\mv{e}$ may not be correctly decoded at the relay. In the case the correction signal is not correctly decoded, the signal $(\mv{b_A}-\mv{e})$ is not  a codeword of the codebook at the lattice level $L_A$, and thus the effective received signal $\mv{x}_{A,PNC}+\mv{x}_{B,PNC}-\phi^{L_A-1}\mv{e}$ is not a  lattice\footnote{In this case, we still name the decoder  as the ``lattice decoder''.}. In particular, the first $L_A-1$ levels of the effective received signals are nested with each other, but they are not nested with the rest of levels of the effective received signals. In this case, the analysis of the decoding FER and coding gain is difficult since the structure of the received signals is quite complicated. We leave this as a future work to further explore. 

\begin{table*}[t]\centering
	\caption{ Rate Setup When $K_B$=537 bits}\label{T_A1}
	\begin{tabular}{|c|c|c|c|c|c|c|c|c|c|}
		\hline
		Scheme                                                                            & \begin{tabular}[c]{@{}c@{}}Time \\ Slot\end{tabular} & User  & \begin{tabular}[c]{@{}c@{}}Modulation \\ Order\end{tabular} & EM Scheme & $R_1$            & $R_2$               & $R_3$               & $R_4$               & \begin{tabular}[c]{@{}c@{}}Overall Rate in \\ This Time Slot\end{tabular} \\ \hline\hline
		
		\multirow{3}{*}{\begin{tabular}[c]{@{}c@{}}ALEM with \\ $K_B$=537 bits\end{tabular}} & \multirow{2}{*}{1}                                   & $A$     & 3                                                         & Lattice  & 0.003            & 0.45             & 0.65             & \textbackslash{} & 0.37                                                                      \\ \cline{3-10}  
		&                                                      & $B$     & 4                                                        & Lattice   & 0.003            & 0.45             & 0.65             & 1                & 0.53                                                                      \\ \cline{2-10} 
		& 2                                                    & $Relay$ & 3                                                        & Lattice   & 0.003            & 0.45             & 0.65             & \textbackslash{} & 0.37                                                                      \\ \hline\hline
		\multirow{5}{*}{\begin{tabular}[c]{@{}c@{}}SLEM with \\ $K_B$=537 bits\end{tabular}} & \multirow{2}{*}{1}                                   & $A$     & 3                                                   & Lattice        & 0.003            & 0.45             & 0.65             & \textbackslash{} & 0.37                                                                      \\ \cline{3-10} 
		&                                                      & $B$     & 3                                                         & Lattice  & 0.003            & 0.45             & 0.65             & \textbackslash{} & 0.37                                                                      \\ \cline{2-10} 
		& 2                                                    & $Relay$ & 3                                                        & Lattice   & 0.003            & 0.45             & 0.65             & \textbackslash{} & 0.37                                                                      \\ \cline{2-10} 
		& 3                                                    & $B$     & 4                                                         & Lattice  & 0.003            & 0.45             & 0.65             & 1                & 0.53                                                                      \\ \cline{2-10} 
		& 4                                                    & $Relay$ & 3                                                          & Lattice & 0.003            & 0.45             & 0.65             & \textbackslash{} & 0.37                                                                      \\ \hline\hline
		\multirow{5}{*}{\begin{tabular}[c]{@{}c@{}}STEM with \\ $K_B$=537 bits\end{tabular}} & \multirow{2}{*}{1}                                   & $A$     & 3                                                     & Polar codes and 8QAM      & \textbackslash{} & \textbackslash{} & \textbackslash{} & \textbackslash{} & 0.37                                                                      \\ \cline{3-10} 
		&                                                      & $B$     & 3                                                     & Polar codes and 8QAM      & \textbackslash{} & \textbackslash{} & \textbackslash{} & \textbackslash{} & 0.37                                                                      \\ \cline{2-10} 
		& 2                                                    & $Relay$ & 3                                                     & Polar codes and 8QAM      & \textbackslash{} & \textbackslash{} & \textbackslash{} & \textbackslash{} & 0.37                                                                      \\ \cline{2-10} 
		& 3                                                    & $B$     & 4                                                      & Polar codes and 16QAM     & \textbackslash{} & \textbackslash{} & \textbackslash{} & \textbackslash{} & 0.53                                                                      \\ \cline{2-10} 
		& 4                                                    & $Relay$ & 3                                                      & Polar codes and 8QAM     & \textbackslash{} & \textbackslash{} & \textbackslash{} & \textbackslash{} & 0.37                                                                      \\ \hline
	\end{tabular}
\end{table*}

\section{Numerical Results}\label{sec:nu}
In this section, we evaluate performance of the  transmission schemes introduced in Section \ref{sec:asy}. In the simulations, we assume that the modulation orders $M_A=3$, and $M_B=4$, i.e., the channel $h_A$ can support user $A$ to transmit signals with modulation order of 3, and the channel $h_B$ can support user $B$ to transmit signals with modulation order of 4. In addition, as an example, we apply the polar codes in the asymmetric and symmetric schemes for data protection. The results of this section can be extended to the case where other channel  codes are applied. We omit the details here.

For the proposed asymmetric transmission scheme, we apply the lattice-based EM shown in Section \ref{sec:en-de},  for uplink and downlink PNC transmissions, and for the correction signal $\hat{\mv{e}}$ transmission. Note that  the downlink PNC transmission and the correction signal transmission are  simple P2P transmissions, and the lattice-based encoding and decoding algorithms shown in Section \ref{sec:en-de} can be similarly applied.  We denote this scheme by ``\textbf{Asymmetric transmission  applying  Lattice-based EM (ALEM)}''.   We have the following two benchmarks:
\begin{itemize}
	\item \textbf{Symmetric transmission  applying Lattice-based EM (SLEM):}
	We apply the lattice-based EM both for uplink and downlink transmissions in time slots 1 to 4. Specifically, in time slot 1, i.e., the uplink of  PNC transmission, the lattice encoding and decoding processes can apply the algorithm shown in Section \ref{sec:lattice} by setting $L_A=L_B$ additionally.  In addition, the transmissions in time slots 2-4 are simple P2P transmissions, and the lattice-based encoding and decoding algorithms shown in Section \ref{sec:en-de} can be applied similarly. 
	\item \textbf{Symmetric transmission applying Traditional EM (STEM):} Unlike SLEM above, in this symmetric transmission scheme here, we apply a traditional EM in which the channel encoding and modulation are two separate processes, instead of the joint encoding and modulation as in the lattice-based EM. In addition, in the uplink of PNC, the relay applies a XOR channel decoder\cite{liew2015primer} to decode the network-coded messages from the two users. 
\end{itemize}

We evaluate the performance in terms of throughput and FER of user $B$. The throughput is defined as follows:
\begin{align}
{\rm Throughput}=\frac{P_BK_B}{T} ~~ bps,
\end{align}
where $P_B$ is the number of times that the whole $K_B$ source  bits are successfully decoded at user $A$ within $T$ duration. Note that, for the symmetric transmission scheme shown in Section \ref{sec:sy}, the $K_B$ information bits are transmitted partially in PNC phase (time slots 1 and 2), and partially in P2P phase (time slots 3 and 4). So the successful decoding of the $K_B$ source bits requires the successful decoding both at time slot 2 and time slot 4 at user $A$. The unit of the throughput is bits per second (bps). In addition, the  bandwidth is 1 $M$ symbols/second. The FER is defined as follows:
\begin{align}
{\rm FER}=1-\frac{P_B}{\bar{P}_B},
\end{align} 
where $\bar{P}_B$ is the number of times that $K_B$ source bits are   decoded at user $A$  within $T$ duration. Moreover, the SNR is defined as 
\begin{align}
SNR=\frac{p_AN}{K_AN_0}~~dB\label{eq:SNR},
\end{align}
where $N_0=2\sigma^2_{R,PNC}$ is the noise power, and $N$ is the packet length in time slot 1 for both schemes. In the simulations, we set $N=256$.

\subsection{Performance When ${\rm mod}_{\phi}\left(\mv{b}_A\right)$ is a Codeword}\label{sec:hy}
\begin{figure}[t]
	\centering
	\includegraphics[scale=0.61]{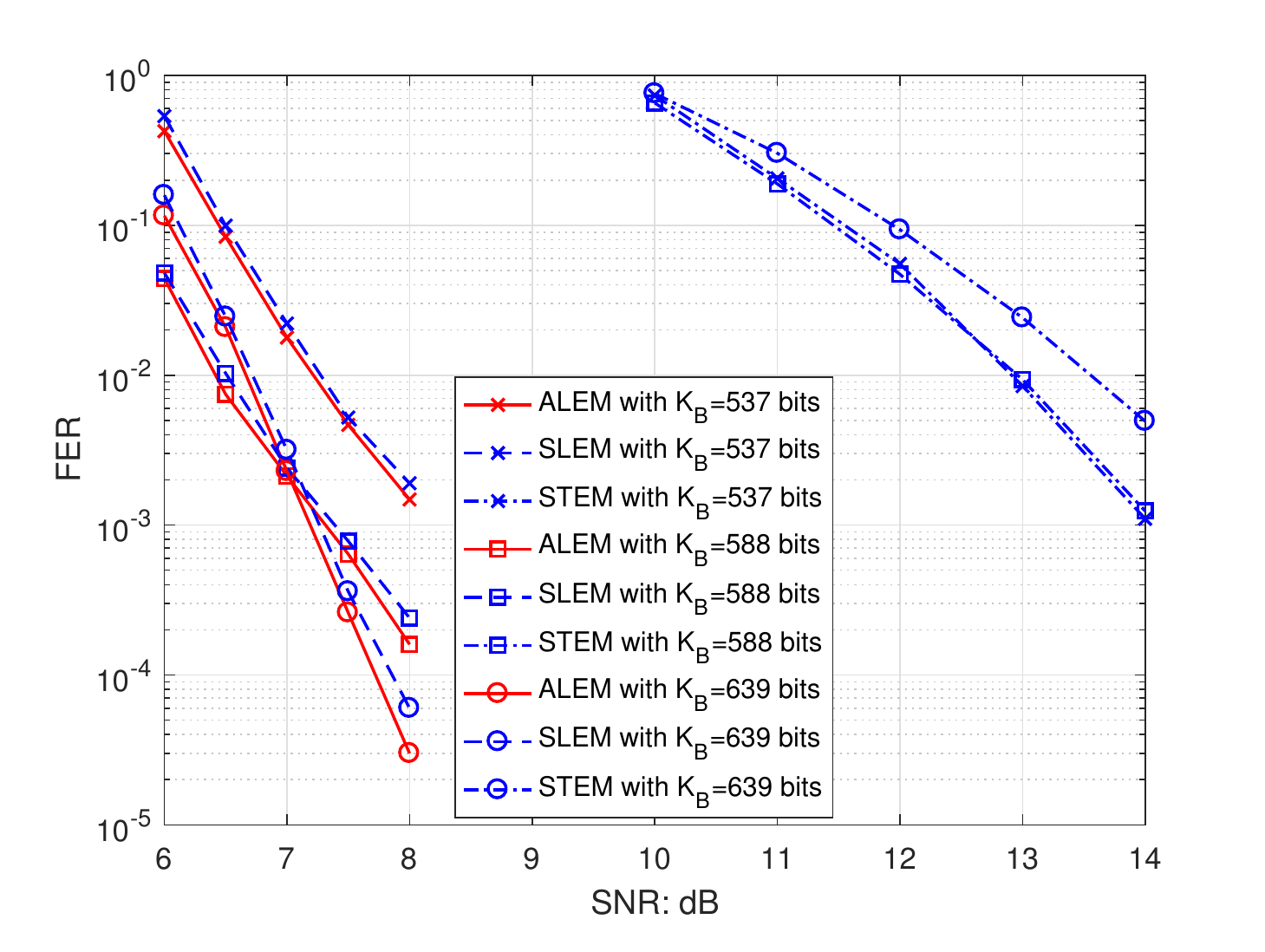}\\
	\caption{FER comparison when ${\rm mod}_{\phi}\left(\mv{b}_A\right)$ is a legal codeword.}\label{Fig6}
\end{figure}

\begin{figure}[t]
	\centering
	\includegraphics[scale=0.61]{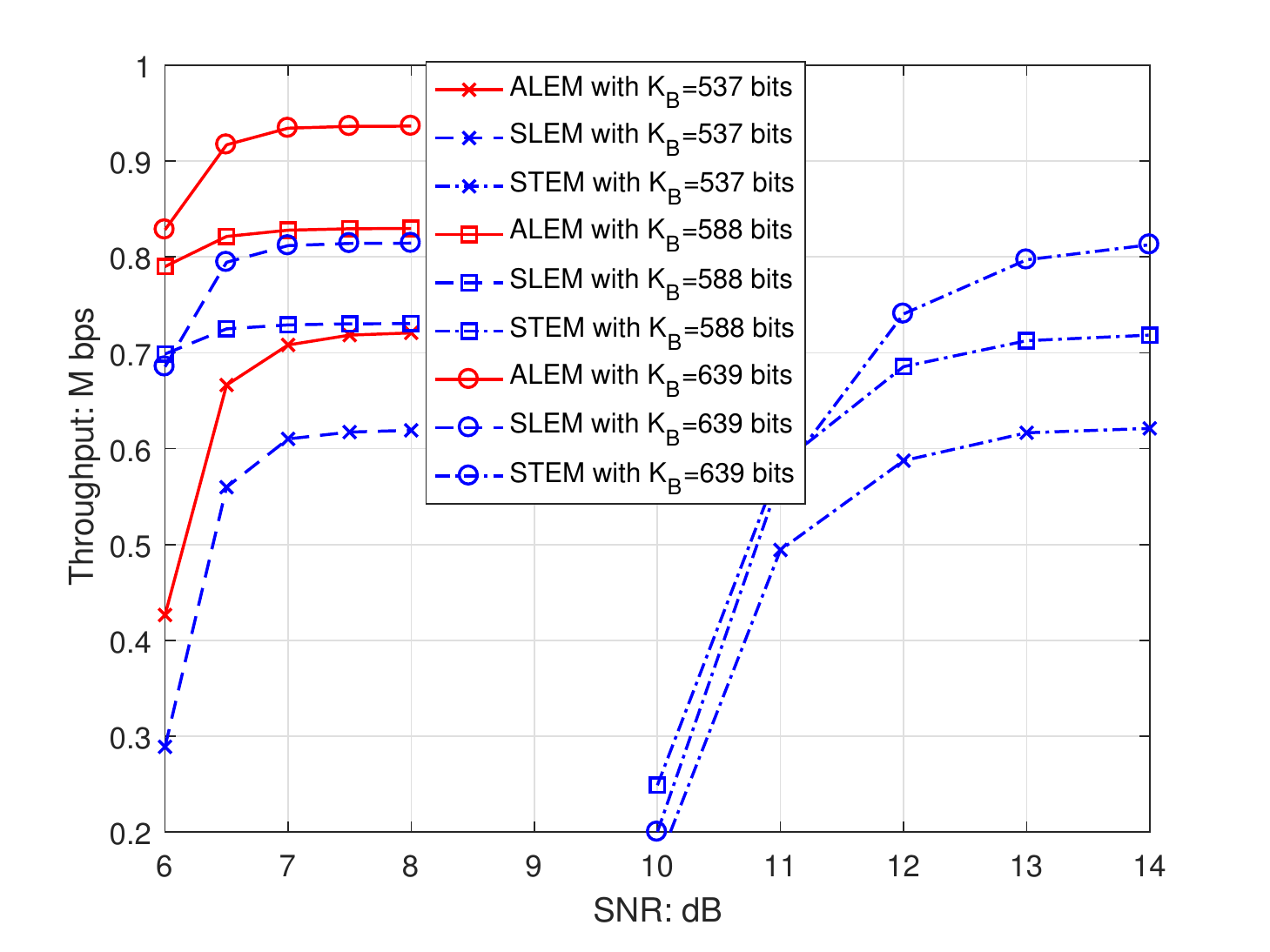}\\
	\caption{Throughput comparison when ${\rm mod}_{\phi}\left(\mv{b}_A\right)$ is a legal codeword.}\label{Fig6_1}
\end{figure}

In the asymmetric transmissions in which $L_A<L_B$, Section \ref{sec:pw_sp}  shows that the hypercube power shaping $\mv{b}_A$ may cause decoding failures at the relay when  ${\rm mod}_{\phi}\left(\mv{b}_A\right)$ is not a codeword to the codebook at the lattice level $L_A$. In this subsection, we first evaluate the performance of the asymmetric transmission in which ${\rm mod}_{\phi}\left(\mv{b}_A\right)$ is a codeword to the codebook at the lattice level $L_A$ by simply setting the rate in lattice level-$L_A$ as 1, i.e., $R_{L_A}=1$. In this case, since the codebook in lattice level-$L_A$ is exactly the space $\{0,1\}^N$,  ${\rm mod}_{\phi}\left(\mv{b}_A\right)$ is a legal codeword to the codebook. As a result, we do not need to transmit the correction signal $\mv{e}$ in the asymmetric transmissions. The results are shown in Figs. \ref{Fig6} and \ref{Fig6_1}.  Let us first introduce the legends in the figures as follows:

\begin{itemize}
	\item \textbf{ALEM with $K_B=537$ bits:} The uplink and downlink rate setups are shown in TABLE \ref{T_A1}. Since $N=256$, user $B$ transmits $K_B=537$ bits in total.  
	\item \textbf{SLEM/STEM with $K_B=537$ bits:} The rate setups of the four time slots are shown in TABLE \ref{T_A1}. Since $N=256$, user $B$ transmits 281 bits in time slot 1. In time slot 3, the user $B$  transmits  256 bits to the relay separately. Thus, user $B$ transmits $K_B=537$ bits in total.   
	\item \textbf{ALEM/SLEM with $K_B=588$ bits:} The setup is the same as ``ALEM/SLEM with $K_B=537$ bits'' except that the coding rate  $R_3=0.85$. 
	\item \textbf{STEM with $K_B=588$ bits:}  The setup is the same as ``STEM with $K_B=537$ bits'' except that the coding rates in time slots 1, 2, 4, are all  $0.43$, and the coding rate in time slot 3 is 0.58. 
	\item \textbf{ALEM/SLEM with $K_B=639$ bits:} The setup is the same as ``ALEM/SLEM with $K_B=537$ bits'' except that the coding rates $R_2=0.55$ and $R_3=0.95$.
	\item \textbf{STEM with $K_B=639$ bits:}  The setup is the same as ``STEM with $K_B=537$ bits'' except that the coding rates in time slots 1, 2, 4, are all  $0.50$, and the coding rate in time slot 3 is 0.63. 
\end{itemize}

We first compare the performance of ALEM and SLEM. First, Fig. \ref{Fig6} shows that for the same amount of the transmitted bits $K_B$, the FERs of ALEM and SLEM are roughly the same. Second, Fig. \ref{Fig6_1} shows that, in terms of throughput, ALEM performs much better than SLEM. For example, when $SNR=8$ dB and $K_B=639$ bits, ALEM has 15\% throughput improvement compared with that of  SLEM. It suggests that, our proposed asymmetric transmission scheme has significant throughput improvement compared with that of the symmetric transmission, since the asymmetric transmission scheme saves the transmission time greatly. We emphasize that the ALEM achieves the above throughput improvement by simply setting the coding rate $R_{L_A}=1$.

Next, we compare the performance of SLEM and STEM. First, from Fig. \ref{Fig6}, for a given FER, SLEM performs 6 dB better than that of STEM. The reasons mainly come from the different PNC decoders at the relay between SLEM and STEM. Specifically, from the decoding process detailed in Section \ref{sec:lattice}, for SLEM the signals fed to the PNC decoder at each level of the lattice  are  BPSK-modulated signals through the modulo $\phi$ operation, while for STEM the signals fed to the PNC detector are 8QAM modulated signals. For a given SNR, the PNC detector  for low-order modulated signals  has better FER performance than that for  high-order modulated signals\cite{liew2015primer}. For example, the detector for BPSK-modulated signals performs much better than the detector for 8QAM-modulated signals in terms of FER. As a result, SLEM performs better than STEM. Second, from Fig. \ref{Fig6_1},  the throughput of SLEM is much higher than that of STEM, since the FER of SLEM is much lower than that of STEM.

\begin{figure}[t]
	\centering
	\includegraphics[scale=0.61]{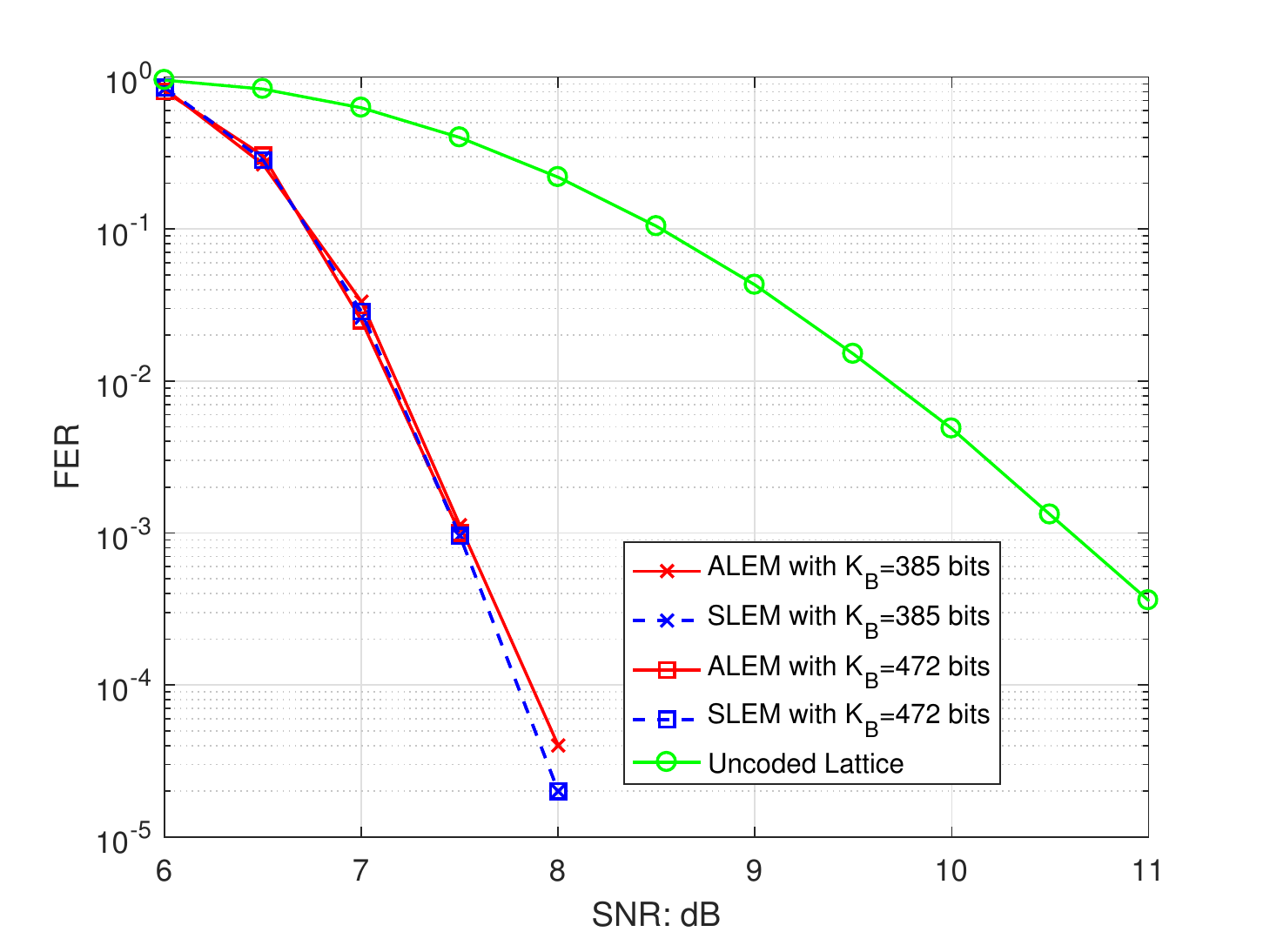}\\
	\caption{FER comparison for general hypercube power shaping.}\label{Fig7}
\end{figure}

\begin{figure}[t]
	\centering
	\includegraphics[scale=0.61]{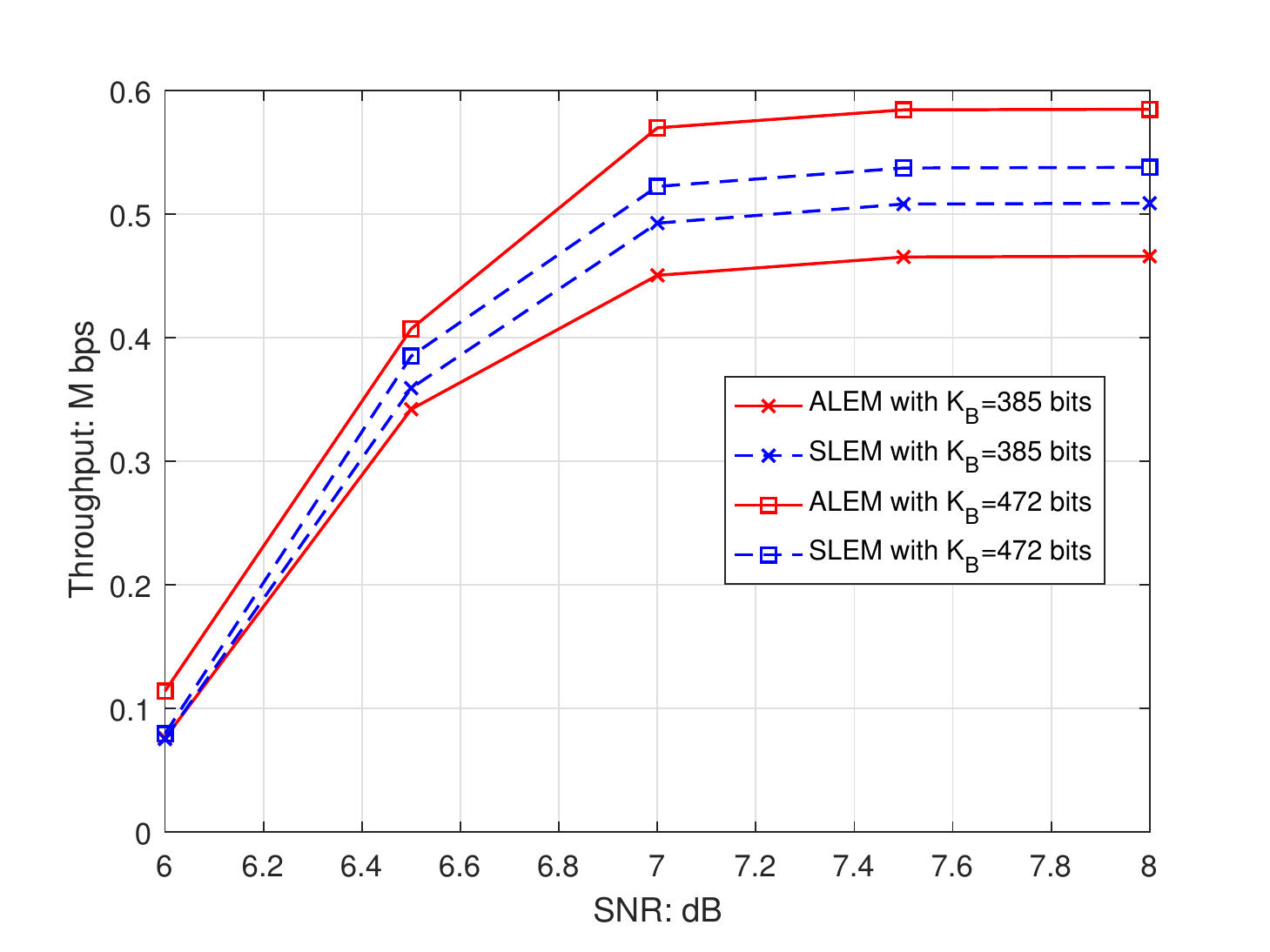}\\
	\caption{Throughput comparison for general hypercube power shaping.}\label{Fig8}
\end{figure}

\begin{figure}[t]
	\centering
	\includegraphics[scale=0.41]{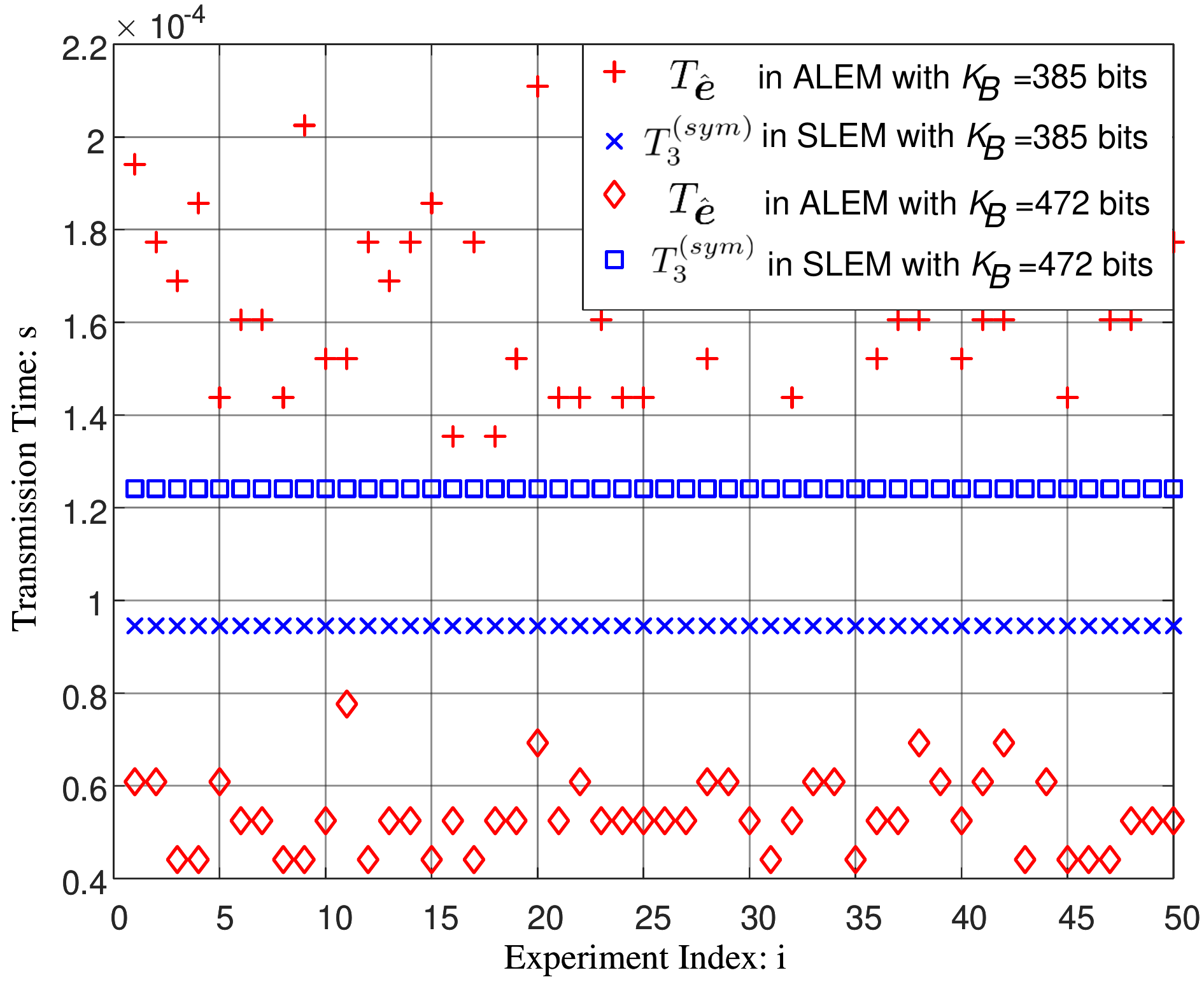}\\
	\caption{$T^{(Sym)}_3$ and $T_{\hat{\mv{e}}}$ comparison over different $K_B$. In ALEM, we plot the time $T_{\hat{\mv{e}}}$ for $K_B=385$ bits and $K_B=472$ bits; in SLEM, we plot the time $T_3^{(sym)}$ for $K_B=385$ bits and $K_B=472$ bits.}\label{Fig9}
\end{figure}

\begin{table*}[]\centering
	\caption{Rate Setup When $K_B$=385 bits}\label{T_B1}
	\vspace{-1em}
	\begin{tabular}{|c|c|c|c|c|c|c|c|}
		\hline
		Scheme                                                                            & \begin{tabular}[c]{@{}c@{}}Time \\ Slot\end{tabular} & User  & \begin{tabular}[c]{@{}c@{}}Modulation \\ Order\end{tabular} & $R_1$             & $R_2$               & $R_3$               & $R_4$               \\ \hline\hline
		\multicolumn{1}{|l|}{\multirow{4}{*}{\begin{tabular}[c]{@{}l@{}}ALEM with \\ $K_B$=385 bits\end{tabular}}} & 0                                                    & $A$     & 3                                                           & 0.003            & 0.40             & 0.55             & \textbackslash{} \\ \cline{2-8} 
		\multicolumn{1}{|l|}{}                                                                                  & \multirow{2}{*}{1}                                   & $A$     & 3                                                           & 0.003            & 0.40             & 0.55             & \textbackslash{} \\ \cline{3-8} 
		\multicolumn{1}{|l|}{}                                                                                  &                                                      & $B$     & 4                                                           & 0.003            & 0.40             & 0.55             & 0.56             \\ \cline{2-8} 
		\multicolumn{1}{|l|}{}                                                                                  & 2                                                    & $Relay$ & 3                                                           & 0.003            & 0.40             & 0.55             & \textbackslash{} \\ \hline\hline
		\multirow{5}{*}{\begin{tabular}[c]{@{}c@{}}SLEM with \\ $K_B$=385 bits\end{tabular}} & \multirow{2}{*}{1}                                   & $A$     & 3                                                           & 0.003            & 0.40             & 0.55             & \textbackslash{}                                                                    \\ \cline{3-8} 
		&                                                      & $B$     & 3                                                           & 0.003            & 0.40             & 0.55             & \textbackslash{}                                                                      \\ \cline{2-8} 
		& 2                                                    & $Relay$ & 3                                                           & 0.003            & 0.40             & 0.55             & \textbackslash{}                                                                      \\ \cline{2-8} 
		& 3                                                    & $B$     & 4                                                           & 0.003            & 0.40             & 0.55             & 0.56                                                                                    \\ \cline{2-8} 
		& 4                                                    & $Relay$ & 3                                                           & 0.003            & 0.40             & 0.55             & \textbackslash{}                                                                       \\ \hline
	\end{tabular}
\end{table*}

\subsection{Performance for General Hypercube Power Shaping}\label{sec:gen}
We next show the simulation results for general hypercube power shaping in Figs.\ref{Fig7}-\ref{Fig9}. Note that, for general hypercube power shaping, ${\rm mod}_{\phi}\left(\mv{b}_A\right)$ may not be a codeword to the codebook at the lattice level $L_A$. In this case, user $A$ may need to send a correction signal to the relay beforehand. Let us first introduce the legends in the figure, as follows:

 \begin{itemize}
	\item \textbf{ALEM with $K_B=385$ bits:}  User $A$ first transmits the compressed correction signal $\hat{\mv{e}}$ to the relay in time slot 0. The rate setup is shown in TABLE \ref{T_B1}. Since  $N=256$,  user $B$ transmits $K_B=385$ bits in total.	
	\item \textbf{SLEM with $K_B=385$ bits:} The rates setup is shown in TABLE \ref{T_B1}. Since $N=256$, user $B$ transmits 68 bits in time slot 1. In time slot 3, the user $B$  transmits  317 bits to the relay separately. Thus, user $B$ transmits $K_B=385$ bits in total.
	\item \textbf{ALEM/SLEM with $K_B=472$ bits:} The setup is the same as ``ALEM/SLEM with $K_B=385$ bits''  except that the coding rate  $R_4=0.90$. 
\end{itemize}

\begin{figure}[ht]
	\centering
	\includegraphics[scale=0.61]{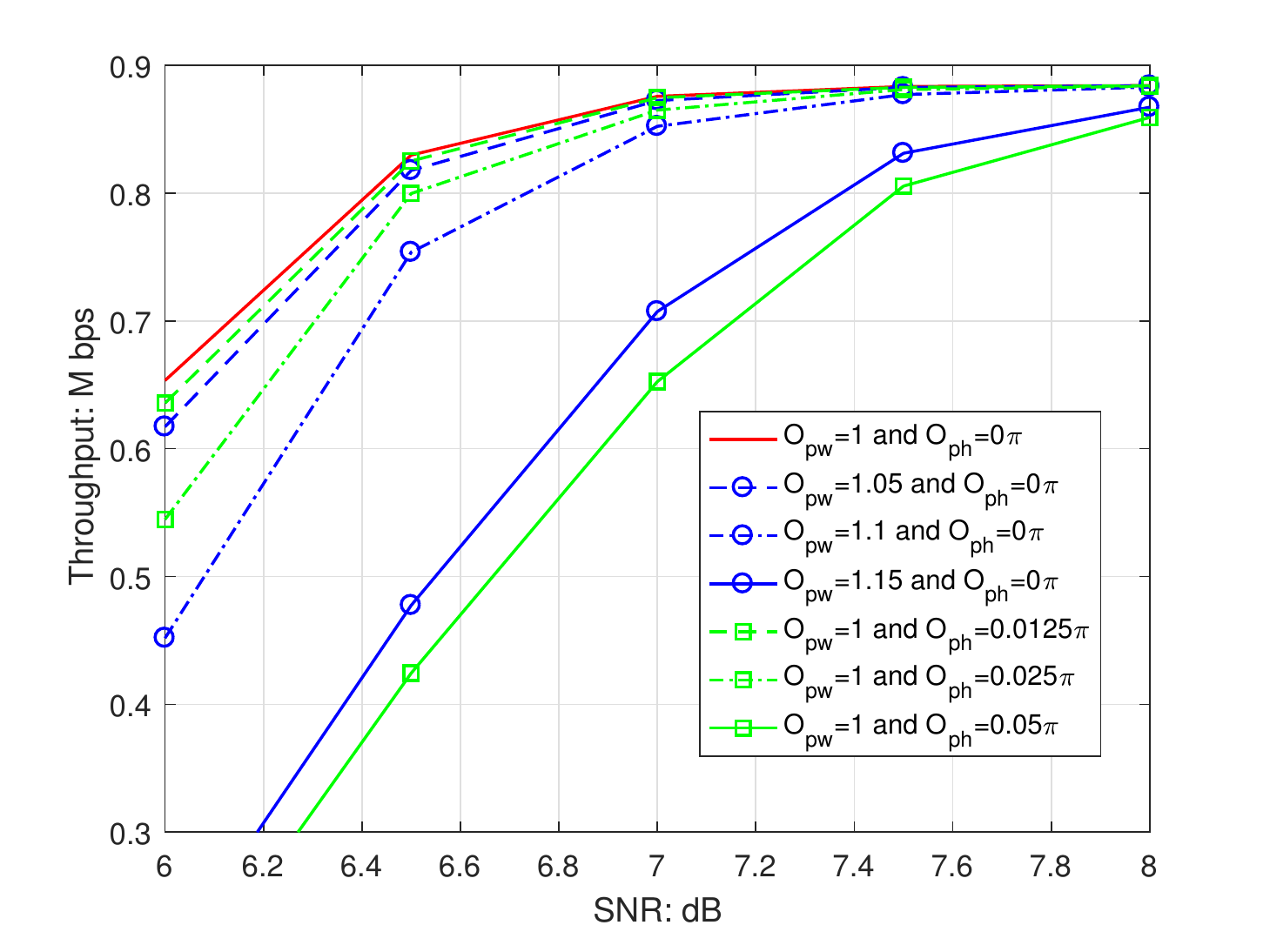}\\
	\caption{ Throughput penalty when the channel precoding is not perfect.}\label{Fig11}
\end{figure}

First, from Fig. \ref{Fig7} we obverse that ALEM and SLEM perform relative the same in terms of FER. Second, Fig. \ref{Fig8} shows that in terms of throughput, when $K_B=472$ bits ALEM outperforms SLEM, while when $K_B=385$ bits SLEM outperforms  ALEM. Since the FERs of the two schemes are roughly the same, this phenomenon is due to the transmission  time $T^{(Sym)}$ of SLEM, and the transmission time $\hat{T}^{(Asy)}$ of ALEM. Specifically, according to \eqref{eq:com}, we have $T^{(Sym)}-\hat{T}^{(Asy)}=T^{(Sym)}_3-T_{\hat{\mv{e}}}$. In this case, in Fig. \ref{Fig9} we compare the correction signal transmission time   $T_{\hat{\mv{e}}}$  and the duration of the third time slot $T^{(Sym)}_3$  over different $K_B$ for 50 Monte Carlo simulations. When $K_B=472$ bits, we observe from Fig. \ref{Fig9} that $T_{\hat{\mv{e}}}<T^{(Sym)}_3$. As a result, from \eqref{eq:com} we have $\hat{T}^{(Asy)}<T^{(Sym)}$. In this case, for a relatively same FER, the throughput of ALEM is higher than that of SLEM; on the other hand,  when $K_B=385$ bits, we observe from Fig. \ref{Fig9} that  $T_{\hat{\mv{e}}}>T^{(Sym)}_3$. As a result, from \eqref{eq:com} we have $\hat{T}^{(Asy)}>T^{(Sym)}$. In this case, for a relatively same FER, the throughput of  SLEM is higher than that of ALEM. In Section \ref{sec:dy}, we will solve this problem by proposing a dynamic transmission scheme.  In addition, from Fig. \ref{Fig7}, benchmarked against the uncoded lattice, we can see that the ALEM has 3dB coding gain at $10^{-3}$ when the correction signal error is taken into account.

\subsection{Performance Penalty for Imperfect Channel Precoding}

\begin{table*}[h]\centering
	\caption{Rate Setup When $K_B$=614 bits}\label{T_B3}
	\vspace{-1em}
	\begin{tabular}{|c|c|c|c|c|c|c|c|}
		\hline
		Scheme                                                                            & \begin{tabular}[c]{@{}c@{}}Time \\ Slot\end{tabular} & User  & \begin{tabular}[c]{@{}c@{}}Modulation \\ Order\end{tabular} & $R_1$             & $R_2$               & $R_3$               & $R_4$               \\ \hline\hline
		\multirow{3}{*}{\begin{tabular}[c]{@{}c@{}}ALEM with \\ $K_B$=614 bits\end{tabular}} & \multirow{2}{*}{1}                                   & $A$     & 3                                                         & 0.003            & 0.45             & 0.95             & \textbackslash{}                                                                      \\ \cline{3-8}  
		&                                                      & $B$     & 4                                                           & 0.003            & 0.45             & 0.95             & 1                                                                                    \\ \cline{2-8} 
		& 2                                                    & $Relay$ & 3                                                          & 0.003            & 0.45             & 0.95             & \textbackslash{}                                                                      \\ \hline
	\end{tabular}
\end{table*}

In this subsection, we evaluate the performance when the channel precoding conditions in \eqref{eq:c1} and \eqref{eq:c2} are not strictly satisfied. We define the channel gain compensation offset as:
\begin{align}
O_{pw}=\frac{|h_B||\beta_B|\sqrt{p_A}}{|h_A||\beta_A|\sqrt{p_B}},
\end{align}
and the channel phase compensation offset as:
\begin{align}
O_{ph}=\theta_{\beta_B}+\theta_{h_B}-\theta_{\beta_A}-\theta_{h_A}.
\end{align}
Note that, in PNC, only $O_{pw}$ and $O_{ph}$ affect the performance, not the individual $\frac{|h_u||\beta_u|}{\sqrt{p_u}}$ and $\theta_{\beta_u}+\theta_{h_u}$, $u\in\{A,B\}$. In Fig. \ref{Fig11}, we show the throughput performance of ALEM under different $O_{pw}$ and $O_{ph}$. The parameters of  ALEM is shown in TABLE \ref{T_B3}. First, when $O_{pw}\le1.1$ or $O_{ph}\le0.025\pi$, we observe from Fig. \ref{Fig11} that the performance penalty is quite small. Second, when $O_{pw}=1.15$ or $O_{ph}=0.05\pi$, Fig. \ref{Fig11} shows that the throughput penalty is 1dB when SNR is low, and the penalty can be ignored when $SNR=8$ dB. We emphasize that, the tested power offsets and phase offsets are based on the channel precoding precision in a lattice-based PNC implementation paper\cite{tan2018mobile}.

\begin{figure}[t]
	\centering
	\includegraphics[scale=0.61]{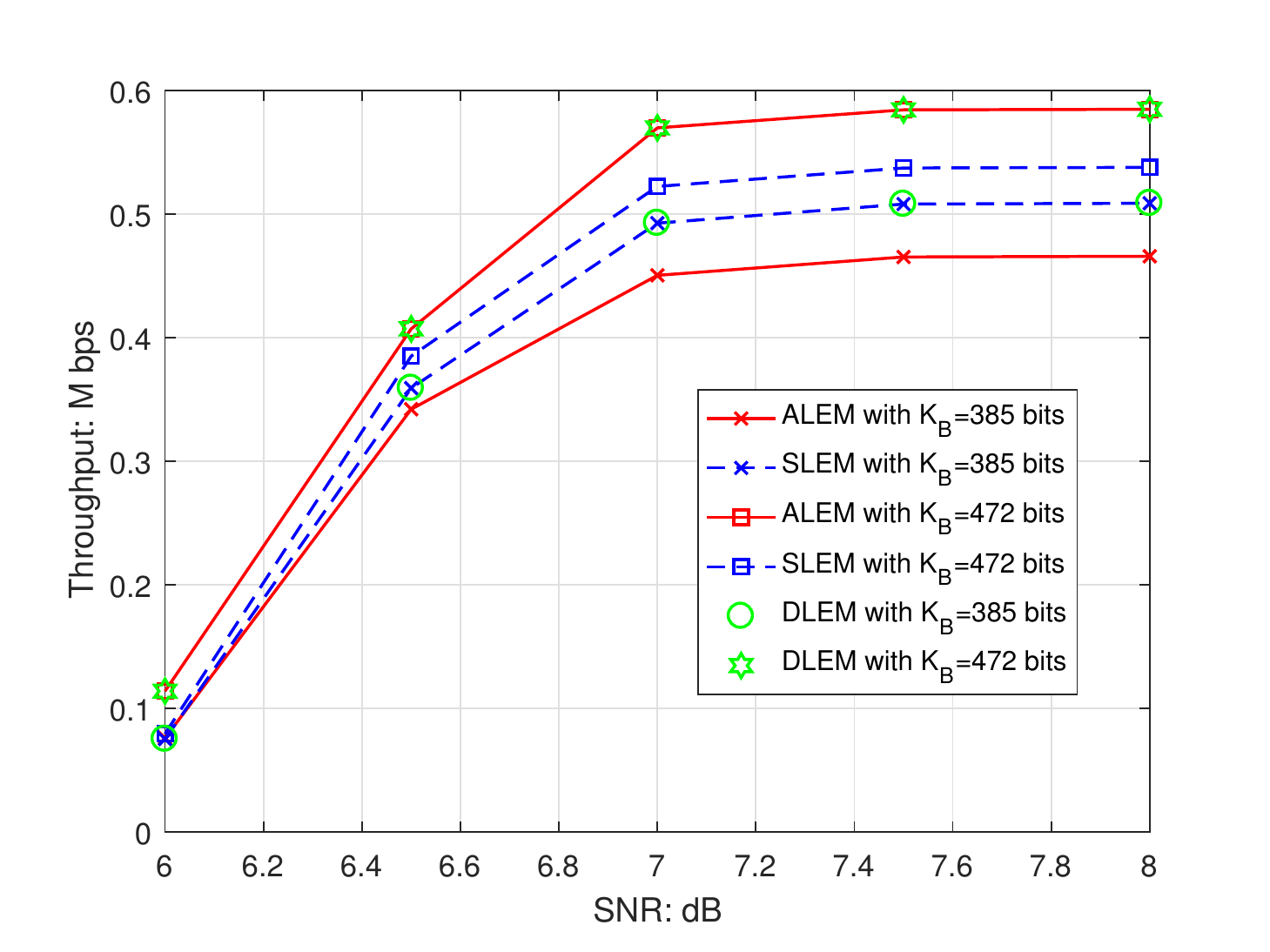}\\
	\caption{Throughput comparison between DLEM, ALEM, and SLEM for general hypercube power shaping.}\label{Fig10}
\end{figure}

\section{Dynamic Transmission Scheme}\label{sec:dy}
As we shown in Section \ref{sec:gen}, when the coding rate $R_{L_A}$ is far smaller than 1, the symmetric transmission may perform better than that of the asymmetric transmission since  $T_{\hat{\mv{e}}}$ may be larger than $T^{(Sym)}_3$. In this case, if we insist on the asymmetric PNC transmission at all, we will not achieve the optimal throughput in the end since sometimes the symmetric transmission scheme performs better. To solve the above problem, we propose to ask the relay $R$ to choose the two schemes dynamically based on the values of $T_{\hat{\mv{e}}}$ and $T^{(Sym)}_3$ so that the dynamic transmission scheme can always achieve the better performance between  ALEM and SLEM. Specifically, if $T^{(Sym)}_3\le T_{\hat{\mv{e}}}$, then the symmetric transmission scheme is selected; otherwise, the asymmetric transmission scheme is selected. We denote this scheme by ``\textbf{Dynamic transmission scheme applying Lattice-based EM (DLEM)}''. The simulation results are shown in Fig. \ref{Fig10}. Let us first introduce the legends as follows.
\begin{itemize}
	\item \textbf{DLEM with $K_B=385/472$ bits:} The relay dynamic selects the scheme between ALEM and SLEM  according to the  values $T_{\hat{\mv{e}}}$ and $T_3^{(Sym)}$. The parameters of ALEM/SLEM with $K_B=385/472$ bits are introduced in Section \ref{sec:gen}.
\end{itemize}
Fig. \ref{Fig10} shows the throughput comparison between DLEM and ALEM/SLEM. From Fig. \ref{Fig10}, DLEM can always achieve the better performance between SLEM and ALEM  over different rate setups, since DLEM always chooses the scheme with smaller transmission time.

\section{Conclusion}\label{sec:con}
This paper studied an asymmetric transmission scheme in PNC in which users $A$ and $B$ transmit different amount of information in the uplink of PNC simultaneously.  A key challenge is how to implement channel coding and modulation in asymmetric PNC transmission  such that the relay can deduce network-coded messages correctly from the two users. To solve this problem, we first proposed a lattice-based EM in which the two users encode and modulate their information in lattices with different lattice levels. In addition, we find that the applied hypercube power shaping causes decoding failures at the relay. We solved this problem by asking user $A$ to transmit a correction signal beforehand such that the difference between the power shaping and the correction signal is a legal codeword. Last, to reduce the correction signal transmission time, we applied the polar source coding technique to compress the correction signal, and user $A$ can transmit the compressed correction signal instead. We find that the polar source coding technique can efficiently reduce the correction signal transmission time when the channel coding rate at lattice level $L_A$ is close to 1. However,  when channel coding rate at lattice level $L_A$ is not close to 1, the overall asymmetric transmission time may be larger than the symmetric time. To solve this problem, we put forth a dynamic transmission scheme in which  the relay  dynamically selects one of the transmission schemes which has smaller transmission time. Numerical results demonstrate the effectiveness of the proposed schemes.

\bibliographystyle{IEEEtran}
\bibliography{CIC}

\begin{IEEEbiography}[{\includegraphics[width=1in,height=1.25in,clip,keepaspectratio]{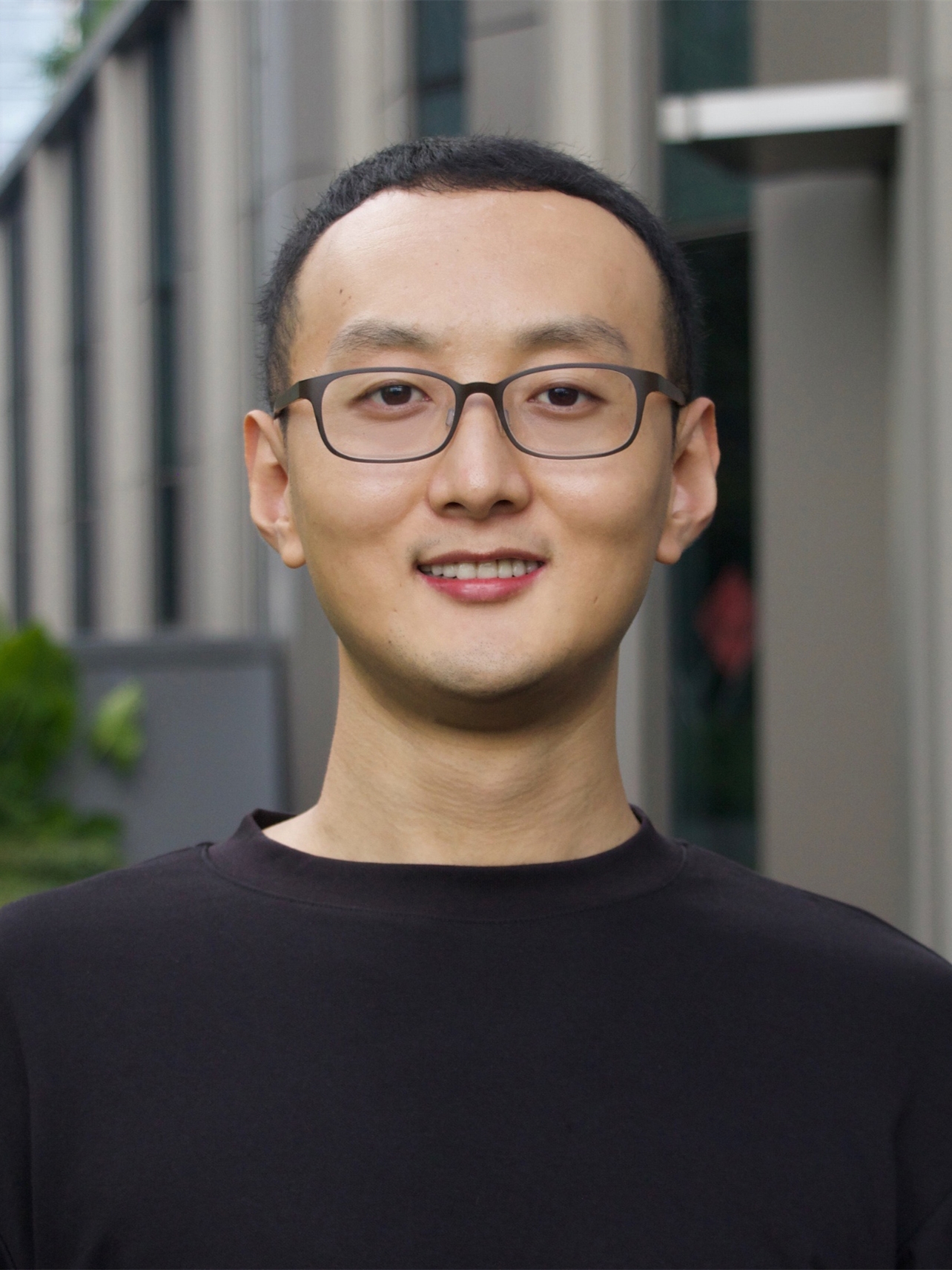}}]{Zhaorui Wang} received the Ph.D. degree from The Chinese University of Hong Kong (CUHK) in 2019, and the B.S. degree from University of Electronic Science and Technology of China in 2015. He was a postdoctoral fellow with the Department of Electronic and Information Engineering, The Hong Kong Polytechnic University, from 2019 to 2020. He is currently a postdoctoral fellow with the Department of Information Engineering at CUHK. He was a recipient of the Hong Kong PhD Fellowship from 2015 to 2018. His research interests include intelligent reflecting surface (IRS) assisted communications, massive machine-type communications (mMTC), and physical-layer network coding (PNC). 
\end{IEEEbiography}

\begin{IEEEbiography}[{\includegraphics[width=1in,height=1.25in,clip,keepaspectratio]{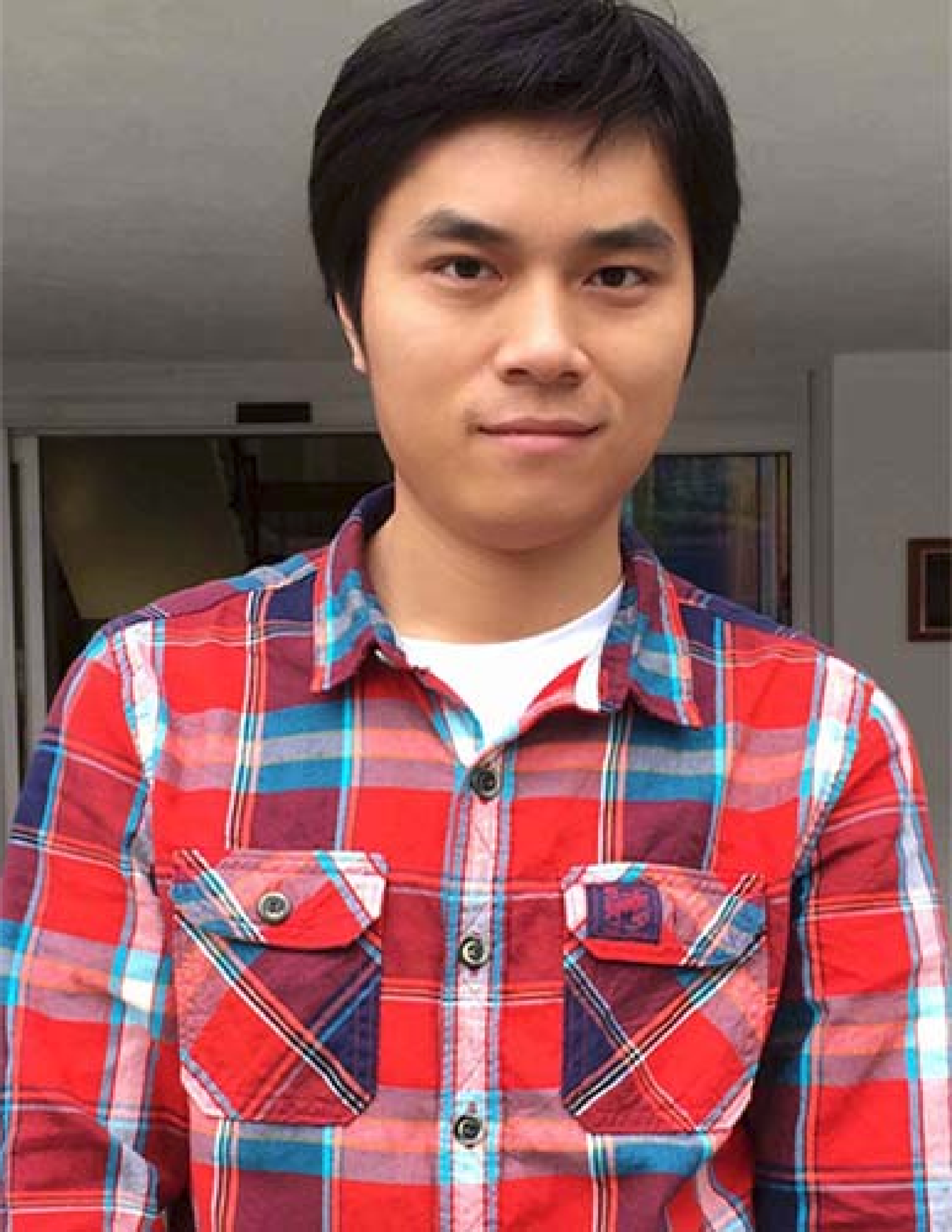}}]{Ling Liu} received the B.S. degree from Nanjing University, Nanjing, China, in 2008, the M.S. degree from Peking University, Beijing, China, in 2012, and the Ph.D. degree from the Department of Electrical and Electronic Engineering, Imperial College London, U.K., in 2016. After that, he was a Research Assistant with CSP Group, Imperial College London. From 2017 to 2019, he served as a Senior Engineer with CT Laboratory, Huawei Technologies, Shenzhen, China. He is currently an Assistant Professor with the Department of Computer Science and Software Engineering, Shenzhen University, Guangdong, China. His research interests are in coding theory, physical layer security, lattice codes, and information theory.
\end{IEEEbiography}

\begin{IEEEbiography}[{\includegraphics[width=1in,height=1.25in,clip,keepaspectratio]{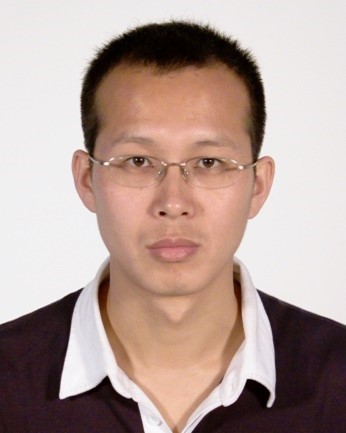}}]{Shengli Zhang} received his B. Eng. degree in electronic engineering and the M. Eng. degree in communication and information engineering from the University of Science and Technology of China (USTC), Hefei, China, in 2002 and 2005, respectively. He received his Ph.D in the Department of Information Engineering, the Chinese University University of Hong Kong (CUHK), in 2008. After that, he joined the Communication Engineering Department, Shenzhen University, where he is a full professor now. From 2014.3 to 2015.3, he was a visiting associate professor at Stanford University. 
	
Shengli Zhang is the pioneer of Physical-layer network coding (PNC). He has published over 20 IEEE top journal papers and ACM top conference papers, including IEEE JSAC, IEEE TWC, IEEE TMC, IEEE TCom and ACM Mobicom. His research interests include blockchain, physical layer network coding, and wireless networks. He is a senior member of IEEE, severed as an editor for IEEE TVT, IEEE WCL and IET Communications. He has also severed as TPC member in several IEEE conferences.
\end{IEEEbiography}

\begin{IEEEbiography}[{\includegraphics[width=1in,height=1.25in,clip,keepaspectratio]{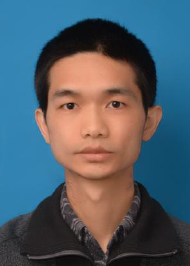}}]{Pengpeng Dong} received the master's degree from University of Science and Technology of China. He is currently a Network Topology and Network Coding Research Expert with Shanghai Huawei Technologies Company, Ltd., Shanghai, China. He is the 5G Network Coding Technologies Research Project Manager. He is the inventor of more than 100 patents. His research interests include network coding, new topology, IIOT communication system, wireless energy harvesting and advanced wireless communication technologies.
\end{IEEEbiography}

\begin{IEEEbiography}[{\includegraphics[width=1in,height=1.25in,clip,keepaspectratio]{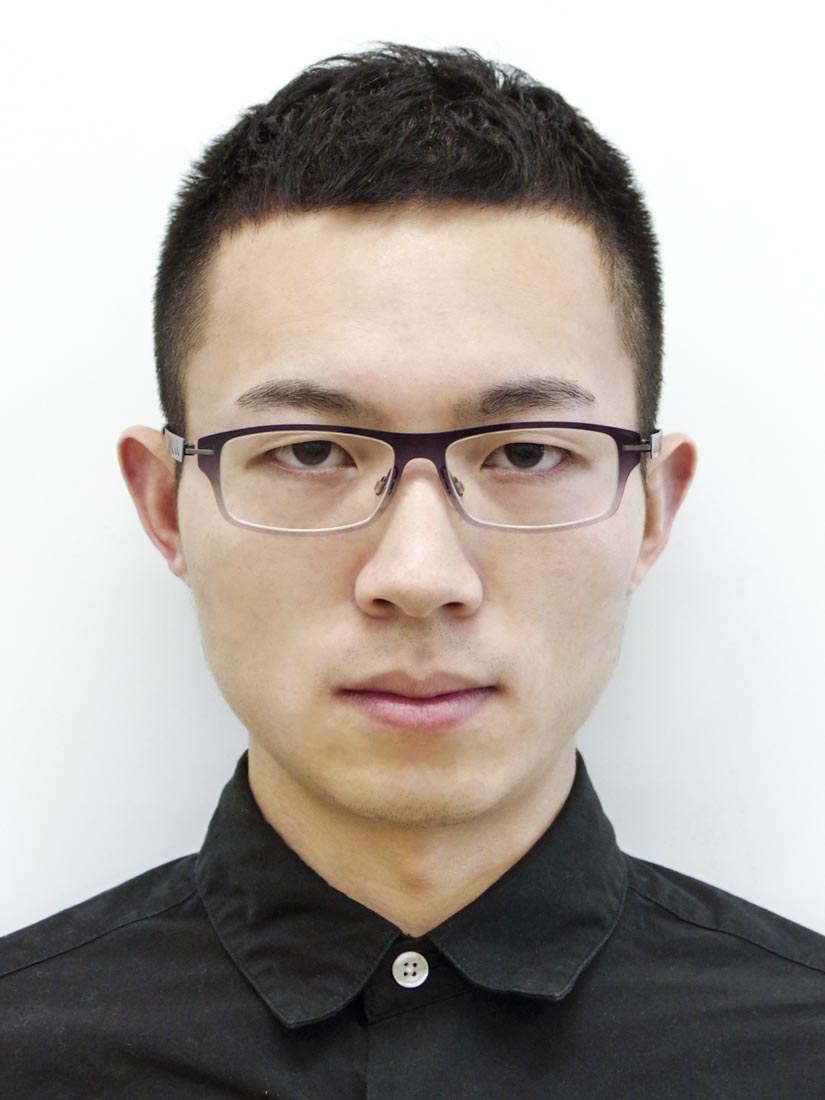}}]{Qing Yang} (M'17) received his B.E. degree (advanced class) from Huazhong University of Science and Technology and Ph.D. degree from The Chinese University of Hong Kong. In 2018, he joined the Shenzhen University as an assistant professor in the College of Electronics and Information Engineering and the principal researcher in the Blockchain Technology Research Center of Shenzhen University. His research interests include blockchain technology, intelligent energy in smart grid, and IoT networking.
\end{IEEEbiography}

\begin{IEEEbiography}[{\includegraphics[width=1in,height=1.25in,clip,keepaspectratio]{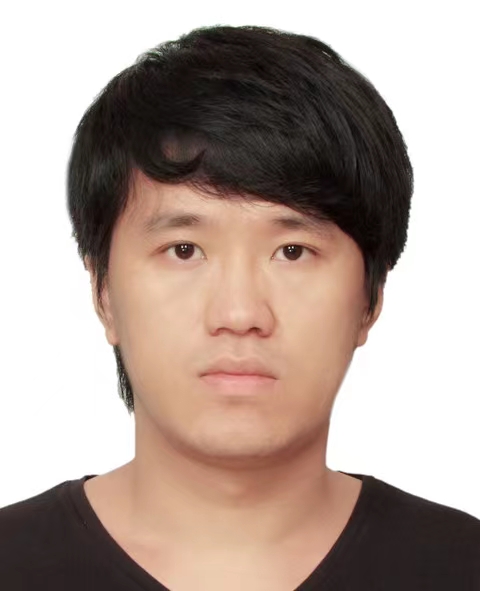}}]{Taotao Wang}(M'16) received the Ph.D. degree in information engineering from The Chinese University of Hong Kong (CUHK), Hong Kong in 2015; the M.S. degree in information and signal processing from the Beijing University of Posts and Telecommunications (BUPT), Beijing, China, in 2011; and the B.S. degree in electrical engineering from the University of Electronic Science and Technology of China (UESTC), Chengdu, China, in 2008. From 2015 to 2016, he was a postdoc research fellow at the Institute of Network Coding (INC) of CUHK. He joined the College of Information Engineering, Shenzhen University as a tenure-track Assistant Professor in 2016 and was promoted as a tenured Associate Professor in 2021. He was served as a TPC member of IEEE CIC ICC 2019, IEEE ICCC 2020, an associate editor of IEEE TENCON 2020, and a symposia chair of IEEE WOCC 2022. He was the recipient of the Hong Kong PhD Fellowship Scheme award in 2011. He received the Star of Tomorrow award at the Microsoft Research Asia in 2012. He received the Excellent Reviewer Award from the journal of IEEE Wireless Communications Letters in 2016. He was the recipient of the Best Paper Award (Runner-up) and the Best Student Paper Award of IEEE CPSCOM 2021.
\end{IEEEbiography}
\end{document}